\documentclass[a4paper,fleqn,usenatbib]{mnras}

\usepackage{newtxtext,newtxmath}
\usepackage{tabularx}
\usepackage{diagbox}
\usepackage[T1]{fontenc}
\usepackage{ae,aecompl}
\usepackage{times}
\usepackage{graphicx}
\usepackage{subfigure}
\usepackage{amsmath}	
\usepackage{amssymb}
\usepackage{aas_macros}
\usepackage{natbib}	
\usepackage{color}
\usepackage{upgreek}
\usepackage{savesym}
\savesymbol{iint}
\savesymbol{iiint}
\usepackage{txfonts}
\restoresymbol{TXF}{iint}
\restoresymbol{TXF}{iiint}
\newcommand{\msun}{\,M$_{\odot}$}

\newcommand{\kms}{\,km~s$^{-1}$}

\newcommand{\ang}{\,$\rm{\AA}$}

\title[The Intermediate-Luminosity Red Transient AT 2017be]{AT~2017be - a new member of the class of Intermediate-Luminosity Red Transients}
\author[Y-Z. Cai et al. ]
{Y-Z. Cai$^{1,2}$\thanks{E-mail: yongzhi.cai@studenti.unipd.it},
A.~Pastorello$^{2}$,
M. Fraser$^{3}$,
M.T. Botticella$^{4}$, 
C.~Gall$^{5}$,
I.~Arcavi $^{6,7,8}$, 
\newauthor 
S. Benetti$^{2}$,
E.~Cappellaro$^{2}$,
N.~Elias-Rosa$^{2}$, 
J.~Harmanen$^{9}$, 
G.~Hosseinzadeh $^{6,7}$,
\newauthor  
D.~A.~Howell $^{6,7}$,
J.~Isern$^{10}$,
T.~Kangas$^{11}$,
E. Kankare$^{12}$,
H.~Kuncarayakti$^{9,13}$,
\newauthor 
P.~Lundqvist$^{14}$,
S.~Mattila$^{9}$,
C.~McCully $^{6,7}$,
T.~M.~Reynolds$^{9}$,
A.~Somero$^{9}$,
\newauthor 
M.~D.~Stritzinger$^{15}$,
G.~Terreran$^{16}$.\\
$^{1}$Universit\`a degli Studi di Padova, Dipartimento di Fisica e Astronomia, Vicolo dell'Osservatorio 2, 35122 Padova, Italy. \\
$^{2}$INAF - Osservatorio Astronomico di Padova, Vicolo dell'Osservatorio 5, 35122 Padova, Italy. \\
$^{3}$School of Physics, O'Brien Centre for Science North, University College Dublin, Belfield, Dublin 4, Ireland\\
$^{4}$INAF-Osservatorio Astronomico di Capodimonte, Salita Moiariello 16, 80131, Napoli, Italy\\
$^{5}$Dark Cosmology Centre, Niels Bohr Institute, University of Copenhagen, Juliane Maries Vej 30, DK-2100 Copenhagen \O, Denmark \\
$^{6}$Department of Physics, University of California, Santa Barbara, CA 93106-9530, USA\\
$^{7}$Las Cumbres Observatory, 6740 Cortona Dr Ste 102, Goleta, CA 93117-5575, USA\\
$^{8}$Einstein Fellow\\
$^{9}$Tuorla Observatory, Department of Physics and Astronomy, University of Turku, V\"ais\"al\"antie 20, FI-21500 Piikki\"o, Finland \\
$^{10}$Institut de Ci\`encies de l'Espai (ICE - CSIC) \& Institut d'Estudis Espacials de Catalunya (IEEC), Campus UAB, 08193 Bellaterra, Barcelona, Spain \\
$^{11}$Space Telescope Science Institute, 3700 San Martin Drive, Baltimore, MD 21218, USA\\
$^{12}$Astrophysics Research Centre, School of Mathematics and Physics, Queen's University Belfast, Belfast BT7 1NN, UK \\ 
$^{13}$Finnish Centre for Astronomy with ESO (FINCA), University of Turku, V\"ais\"al\"antie 20, FI- 21500 Piikki\"o, Finland \\
$^{14}$Department of Astronomy and the Oskar Klein Centre, Stockholm University, AlbaNova, SE-106 91 Stockholm, Sweden\\
$^{15}$Department of Physics and Astronomy, Aarhus University, Ny Munkegade 120, DK-8000 Aarhus C, Denmark\\
$^{16}$Center for Interdisciplinary Exploration and Research in Astrophysics CIERA, Department of Physics and Astronomy, \\Northwestern University, Evanston, IL 60208, USA \\
 }
\date{Accepted 2018 July 25. Received 2018 June 29; in original from 2018 April 13}
\pubyear{2018}

\begin{document}
\label{firstpage}
\pagerange{\pageref{firstpage}--\pageref{lastpage}}
\maketitle

\begin{abstract}   
We report the results of our spectrophotometric monitoring campaign for AT~2017be in NGC~2537. Its lightcurve reveals a fast rise to an optical maximum, followed by a plateau lasting about 30 days, and finally a fast decline. Its absolute peak magnitude ($M_{r}$ $\simeq$ $-$12 $\rm{mag}$) is fainter than that of core-collapse supernovae, and is consistent with those of supernova impostors and other Intermediate-Luminosity Optical Transients. The quasi-bolometric lightcurve peaks at $\sim$ 2 $\times$ 10$^{40}$ erg s$^{-1}$, and the late-time photometry allows us to constrain an ejected $^{56}$Ni mass of $\sim$ 8 $\times$ 10$^{-4}$\msun. The spectra of AT~2017be show minor evolution over the observational period, a relatively blue continuum showing at early phases, which becomes redder with time. A prominent H$\alpha$ emission line always dominates over other Balmer lines. Weak Fe {\sc ii} features, Ca~{\sc ii} H$\&$K and the Ca {\sc ii} NIR triplet are also visible, while P-Cygni absorption troughs are found in a high resolution spectrum. In addition, the [Ca~{\sc ii}] $\lambda$7291,7324 doublet is visible in all spectra. This feature is typical of Intermediate-Luminosity Red Transients (ILRTs), similar to SN~2008S. The relatively shallow archival Spitzer data are not particularly constraining. On the other hand, a non-detection in deeper near-infrared HST images disfavours a massive Luminous Blue Variable eruption as the origin for AT~2017be. As has been suggested for other ILRTs, we propose that AT~2017be is a candidate for a weak electron-capture supernova explosion of a super-asymptotic giant branch star, still embedded in a thick dusty envelope.
\end{abstract}

\begin{keywords}
supernovae: general -- galaxies: individual (NGC~2537) -- supernovae: individual (AT~2017be) -- stars: AGB and super-AGB stars --  stars: mass-loss
\end{keywords}
 
\newpage
\section{Introduction} \label{intro}
In recent years, the proliferation of wide-field sky surveys has led to the discovery of a large variety of transient events of varying luminosities. In particular, an increasing number of events show luminosities intermediate between those of core-collapse (CC) supernovae (SNe) and classical novae \citep[see, e.g.,][for a short review on the subject]{2012PASA...29..482K}. These events are generally called intermediate-luminosity optical transients (ILOTs), and are challenging our observational and theoretical understanding of standard scenarios of stellar evolution. 

A sub-set of ILOTs are designated ``SN impostors'' \citep{2000PASP..112.1532V}, and mimic the observables of real SN explosions, although their progenitor stars survive the outburst. In many cases their peculiarity is not immediately recognised and in fact, several impostors have been labelled in the past with supernova designation. SN impostors have absolute magnitudes fainter than $-$15 mag, and their spectra are dominated by prominent hydrogen lines in emission, which usually lack high-velocity components in their profile \citep{2011MNRAS.415..773S, 2016ApJ...823L..23T}. Various scenarios have been proposed to explain SN impostors, including outbursts triggered by binary interaction, or violent mass-loss events during the late evolution of single massive stars. Stars producing SN impostors range from the most massive luminous blue variables \citep[LBVs; e.g.,][]{1989A&A...217...87W,1994PASP..106.1025H,2004ApJ...615..475S} to potentially lower-mass super-asymptotic giant branch (S-AGB) stars \cite[see, e.g.,][]{ 1980PASJ...32..303M, 1984ApJ...277..791N, 1987ApJ...318..307M,  1993ApJ...414L.105H, 2006A&A...450..345K,  2008ApJ...675..614P}.  

LBVs form a class of massive stars with high luminosity, blue colours, large variability, enormous mass-loss rates, and H-rich spectra. The LBV phase is believed to be a short-duration stage in the evolution of massive stars. \citet{1994PASP..106.1025H} proposed that the transition of LBVs from quiescence to the eruptive phase is characterised by an increase in the optical luminosity accompanied by a modest colour evolution (towards redder colours). The best known examples of LBV eruptions are the `Giant Eruption' of the galactic LBV $\eta$~Carinae in the mid-19th Century \citep{1994PASP..106.1025H}  and P~Cygni in the 17th Century \citep{1969BAN....20..225D, 1988IrAJ...18..163D}. In addition to these eruptions in the Milky Way, similar non-terminal outbursts are also occasionally discovered in extra-Galactic environments \citep{2011MNRAS.415..773S}, where they are the most frequently observed SN impostors. 

An alternative scenario to explain some putative SN impostors is the explosion of a Super Asymptotic Giant Branch (S-AGB) star as an electron capture supernova (ECSN) \citep[see, e.g.,][]{2008ApJ...681L...9P, 2009ApJ...705.1425P, 2009ApJ...705.1364T}. These stars are usually surrounded by a thin helium shell and an extended hydrogen envelope~\citep{1987ApJ...322..206N}, and may experience major mass-loss accompanied by significant outbursts. It has been proposed that when a star's initial mass is in the range $\sim$ 8-12\msun,~ it can die as an ECSN \citep[see, e.g.,][]{1980PASJ...32..303M, 1984ApJ...277..791N, 1987ApJ...322..206N, 1987ApJ...318..307M, 1993ApJ...414L.105H, 2006A&A...450..345K,  2008ApJ...675..614P, 2011ApJ...726L..15W}. The Ne-O cores of these H-exhausted stars are very close to the boundary for a SN explosion. Electron-capture reactions by $^{24}$Mg and $^{20}$Ne nuclei reduce electron pressure, triggering the core to collapse before Ne burning ignites, and giving rise to an ECSN. We note however, that the precise mass range for ECSN progenitors is controversial and is likely limited to $\Delta \mathrm{M} \sim$ 1\msun\ or less \citep{2007A&A...476..893S, 2015MNRAS.446.2599D, 2017PASA...34...56D}. 

Recently, \citet{2012MNRAS.424..855K} proposed that CSM interaction following an ECSN can explain the Type IIn SN 2009kn (peak $M_B \approx -18$ mag), as did \citet{2013MNRAS.434..102S} and \citet{2013MNRAS.431.2599M} for the Type IIn SN 2011ht. However, ECSNe are expected to be fairly faint events \citep[e.g.,][]{2013ApJ...771L..12T}. Therefore, candidates are expected to have modest luminosities and type IIn SN like spectra \citep[e.g.,][]{1990MNRAS.244..269S} if a mass loss event happened a short time before the SN explosion. The faint SN~2008S is an appealing ECSN candidate \citep[see, e.g.,][]{2008CBET.1330....1W,  2008CBET.1340....1Y,  2008CBET.1381....1W,  2008ApJ...681L...9P, 2009ApJ...697L..49S, 2009MNRAS.398.1041B, 2010MNRAS.403..474W, 2011ApJ...741...37K, 2012ApJ...750...77S, 2016AAS...22723706S, 2016MNRAS.460.1645A}. In fact, a possible S-AGB progenitor has been proposed by several authors \citep{ 2009MNRAS.398.1041B, 2012ApJ...750...77S, 2016MNRAS.460.1645A}, based on pre-outburst archival Spitzer images which reveal a candidate in the correct mass range, and which was embedded in a dusty cocoon. SN~2002bu \citep{2002IAUC.7863....1P, 2012ApJ...760...20S}, NGC 300 OT2008-1 \citep[e.g.,][]{2009ApJ...695L.154B, 2009ApJ...699.1850B, 2009ApJ...705.1425P, 2010ApJ...709L..11K, 2010A&A...510A.108P, 2011ApJ...743..118H}, M85 OT2006-1 \citep[see, e.g.,][]{2007Natur.447..458K, 2007Natur.449E...1P, 2007ApJ...659.1536R}, PTF10fqs \citep{2011ApJ...730..134K}, NGC 5775 OT2012-1 \citep{2012ATel.4009....1B, 2012PZ.....32....2B} are also very likely ILOTs arising from ECSNe. These transients are collectively referred to as intermediate-luminosity red transients (ILRTs).

In this article, we present a study of a newly discovered ILRT, AT~2017be (also known as KAIT-17A or iPTF17be). AT~2017be was first discovered on 2017 January 06.508 (UT will be used hereafter) by the Lick Observatory Supernova Search (LOSS), at an unfiltered magnitude of 18.5 \citep{2017TNSTR..33....1S}. We note that \citet{2017TNSCR..43....1H} first classified AT~2017be as an LBV, while \citet{2017arXiv171110501A} recently presented some followup observations of AT~2017be, and also proposed an LBV eruption scenario for this event. This work presents a more extensive dataset, with higher-cadence spectro-photometric coverage, and a monitoring window covering all phases of the transients evolution, from which we reach a somewhat different conclusion as to the nature of AT 2017be.  

This paper is organised as follows: in Section \ref{host}, we characterise the host galaxy NGC 2537. In Section \ref{photometry}, we describe the photometric data reduction techniques and present the lightcurves of AT~2017be. In Section \ref{spectroscopy}, the spectroscopic data are shown and analysed. In Section \ref{progenitor}, we examine the pre-explosion frames, in particular $\textit{Hubble Space Telescope}$ ({\sl HST}) archival images, and characterise the quiescent progenitor star a few years before the AT~2017be event.  A discussion on the nature of AT~2017be and a short summary  are presented in Section \ref{discussion}.  

\section{The host galaxy} \label{host} 
\begin{figure} 
\includegraphics[width=1.0\columnwidth]{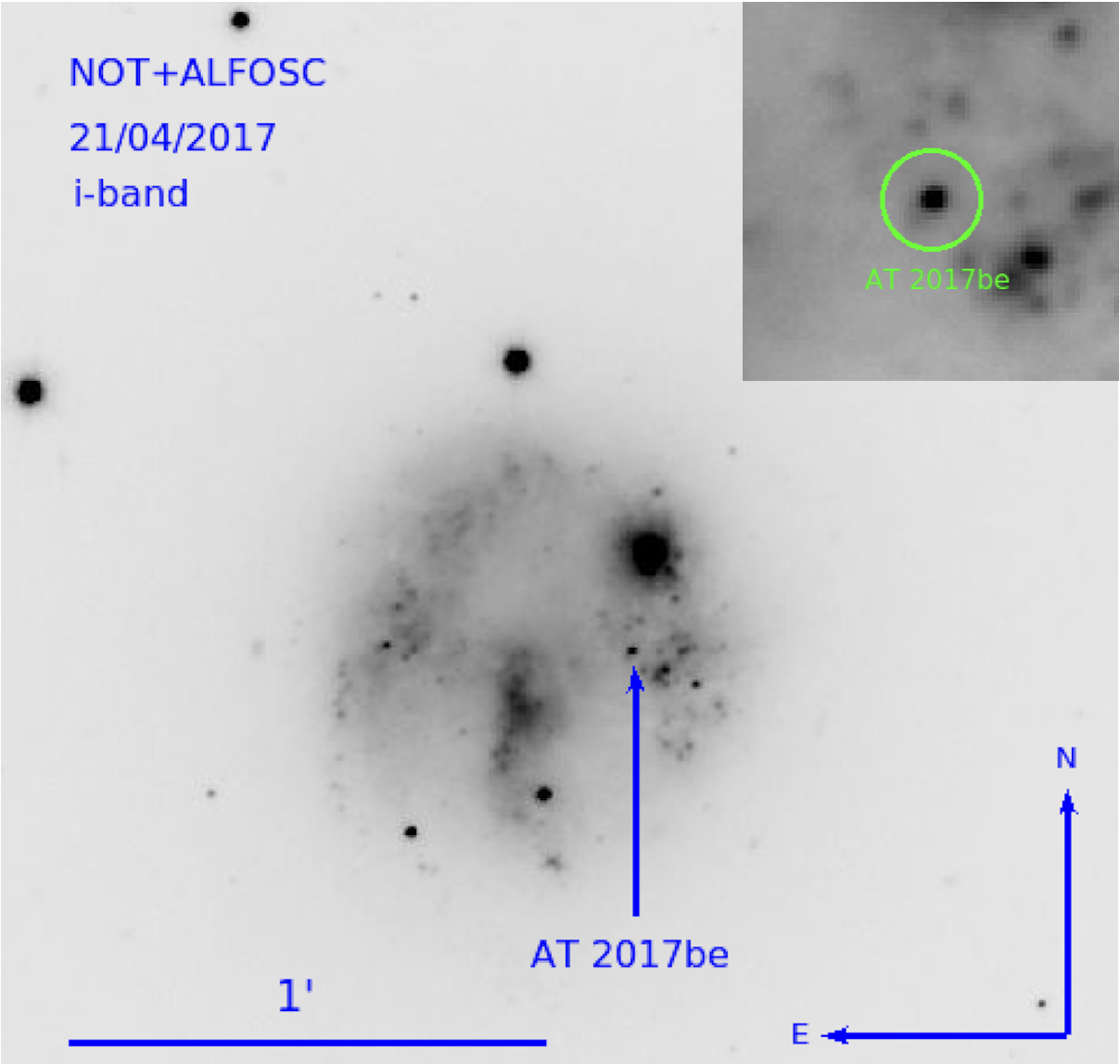}
\caption{The field of AT~2017be. A zoom in on the location of AT 2017be is shown in the inset.}
\label{fc}
\end{figure}

AT~2017be was discovered in NGC~2537 at RA = $08^{h}13^{m}13\fs38$, Dec = +$45\degr59\arcmin29\arcsec.54$~[J2000] (see Figure~\ref{fc}). The host galaxy is also known as Mrk 86, Arp 6 or the Bear's Paw Galaxy.  According to the standard morphological classification scheme \citep{1926CMWCI.324....1H, 1959HDP....53..275D}, NGC 2537 is a blue  SB(rs) galaxy. Detailed information on the host galaxy taken from the NASA/IPAC Extragalactic database\footnote{\url{http://nedwww.ipac.caltech.edu/}} ({\sl NED}) is reported in Table~\ref{galaxy}.  

\begin{table}
\caption{Basic parameters of NGC~2537.}
\label{galaxy}
\begin{tabular}{llc}
\hline
$\alpha$ (J2000) & $08^{h}13^{m}14\fs64$\\					 
$\delta$ (J2000) & $45\degr59\arcmin23\arcsec$3\\
Major Axis (arcmin)  & 1.7\arcmin    \\   
Minor Axis (arcmin)  & 1.5\arcmin    \\
Morphological Type &  SB(rs)dm  \\
Apparent Magnitude (V)&12.32 mag  \\
Redshift & 0.001438 $\pm$ 0.000003	  \\
$v_\mathrm{Heliocentric}$  & 431 $\pm$ 1 \kms   \\
Adopted Distance & 7.82 $\pm$ 0.54 Mpc \\
Galactic Absorption & $A_B$ = 0.195 mag \\
\hline
\end{tabular}
\\[1.5ex]

\end{table}
The distance to NGC~2537 has been estimated with multiple methods, including using the brightest stars in the galaxy \citep{1999AstL...25..322S} and the Tully-Fisher technique \citep{1984A&AS...56..381B, 1988NBGC.C....0000T}. However, the reported estimates are largely discrepant. In particular, different estimates based on the Tully-Fisher method give a wide range of values ($\sim$~9 - 20 $\rm{Mpc}$). From the  radial velocity ($cz$) corrected for local group infall into Virgo ($V_{Vir}$ = 571 $\pm$ 4 \kms)~from LEDA\footnote{\url{http://leda.univ-lyon1.fr/}} \citep{2014A&A...570A..13M} and adopting standard cosmological parameters ($H_{0}$ = 73 \kms $\rm{Mpc}^{-1}$, $\Omega_{M}$ = 0.27, $\Omega_{\Lambda}$ = 0.73), we derive a luminosity distance $d_0$ = 7.82 $\pm$ 0.54 $\rm{Mpc}$ \footnote{This value is obtained using Ned Wright's Cosmological Calculator \url{http://www.astro.ucla.edu/~wright/CosmoCalc.html/}; see \citet{2006PASP..118.1711W}}. The resulting distance modulus $\mu$ = 29.47 $\pm$ 0.15 $\rm{mag}$ will be used throughout this paper. We note that this kinematic distance is in good agreement with that inferred from the absolute magnitude of the brightest stars in NGC~2537 \citep{1999AstL...25..322S}.

We adopt a Milky Way reddening \hbox{E($B-V$)$_\rmn{Gal}$ = 0.048}~mag \citep[from][]{2011ApJ...737..103S} towards AT~2017be. A narrow Na~{\sc i} D absorption is visible in our spectra at the redshift of the host galaxy NGC~2537, but with low signal-to-noise (S/N) ratio. We increased the S/N by averaging the early spectra, and measured for the blended doublet an equivalent width (EW) of $\simeq$ 0.57 $\pm$ 0.17~\AA. Following the relation in \citealt{Turatto 2003} (i.e., E($B-V$) = 0.16 $\times$ EW(Na I D)), and assuming $R_V$ = 3.1 \citep{1989ApJ...345..245C}, we obtain a host galaxy reddening value of \mbox{E($B-V$)$_\rmn{host}$ = 0.04 $\pm$ 0.02~mag} and a total line-of-sight colour excess \mbox{E($B-V$)$_\rmn{total}$ = 0.09 $\pm$ 0.03~mag}. 

Our highest resolution spectrum (+30.2 d from the $r$-band maximum, see Section \ref{spectroscopy}) allows us to deblend the narrow Galactic and host galaxy Na ID absorption line components, which appear to have similar strengths. This supports our claim of a similar reddening contribution from the the Milky Way and NGC~2537. A final check has been performed by computing the Balmer decrement (which is the ratio of the H$\alpha$ and H$\beta$ line intensities) in our spectra. We obtained decrements roughly in agreement with the classical value of H$\alpha$/H$\beta$ = 2.86 for typical gas conditions \citep{1989SvA....33..694O} adopting an E($B-V$)$_\rmn{total}$ = 0.09~mag. This total reddening estimate will be used throughout the paper.


\section{Photometry} \label{photometry}
\subsection{Observations and data reduction} \label{datareduction}
As mentioned previously, the transient AT~2017be was first found by the Lick Observatory Supernova Search (LOSS; reported in {\sl TNS}\footnote{\url{https://wis-tns.weizmann.ac.il/object/2017be}}). An early-phase photometric measurement was also obtained by the Intermediate Palomar Transient Factory (iPTF\footnote{\url{https://www.ptf.caltech.edu/iptf/}}). We included these data in our lightcurve. Our multi-band monitoring campaign started on  2017 January 6, soon after the classification, when we triggered systematic follow-up observations of the target, with a more aggressive strategy during the first month, when the evolution of the transient was more rapid. Subsequently, we slightly relaxed the cadence of the campaign, which went on to span a period of approximately one year. Detailed information on the observations including the photometric measurements are provided in Table~\ref{optical_bands}. 

The optical follow-up campaign of AT~2017be was performed using the following instruments: the 1-m telescope of the Las Cumbres Observatory (LCO\footnote{LCO is a global robotic network of telescopes; see \url{http://lco.global/}}) at McDonald Observatory (Texas, USA) equipped with a Sinistro (labelled in this paper as fl05) camera; the 2.56-m Nordic Optical Telescope (NOT\footnote{\url{http://www.not.iac.es/}}) equipped with ALFOSC and the 10.4-m Gran Telescopio Canarias (GTC\footnote{\url{http://www.gtc.iac.es/}}) with OSIRIS, both located at Roque de los Muchachos Observatory (La Palma, Canary Islands, Spain); and the 1.82-m Copernico Telescope of the INAF - Padova Observatory  at Mt. Ekar (Asiago, Italy)\footnote{\url{http://www.oapd.inaf.it/}} with the AFOSC camera. A complementary near-infrared (NIR) photometric campaign started on 2017 February 7 (approximately one month after the discovery) and ended on 2018 January 4. The NIR data were obtained using the NOT equipped with NOTCam.

All optical imaging frames were  first pre-processed using standard tasks in \textsc{iraf}\footnote{\url{http://iraf.noao.edu/}} for overscan, bias and flat field corrections, and trimming of unexposed frame regions. The steps necessary to obtain the SN magnitude were performed using  the dedicated pipeline {\sc SNoOpy}\footnote{\url{http://sngroup.oapd.inaf.it/snoopy.html/}} developed by \citet{snoopyref}. {\sc SNoOpy} consists of a collection of PYTHON-based scripts calling standard {\sc iraf} tasks within {\sc pyraf}. The package also makes use of other public tools, including {\sc sextractor}\footnote{\url{www.astromatic.net/software/sextractor/}} for source extraction from images, {\sc daophot}\footnote{\url{http://www.star.bris.ac.uk/~mbt/daophot/}} \citep[e.g.,][]{1987PASP...99..191S}~for instrumental magnitude measurements using the point-spread function (PSF) fitting technique, and {\sc hotpants}\footnote{\url{http://www.astro.washington.edu/users/becker/v2.0/hotpants.html/}} \citep[e.g.,][]{2015ascl.soft04004B}~for PSF matching and template subtraction. 

We used the PSF-fitting technique to measure instrumental magnitudes because, with a complex background, it provides more reliable results than aperture photometry. Firstly, we subtracted the underlying background as calculated via a low-order polynomial fit. Then, using the profiles of isolated stars in the  field of AT~2017be, we obtained the PSF model to be used for extracting the instrumental  magnitude of the transient. The measured source is subtracted from the image and a new estimate for local background is obtained. Then the fitting procedure is iterated to improve the results. Error estimates are performed via artificial star experiments, in which fake stars with similar magnitude as the object are positioned very close to the location of AT~2017be. The simulated sources are then measured by the same PSF-fitting techniques, and the dispersion of these measurements provides an estimate of the recovered magnitude errors. However at late phases, when the transient became too faint to obtain reliable measurements with the PSF-fitting technique, we used template subtraction to properly remove the flux contamination of the host galaxy. The template images of the host galaxy were observed as part of the Sloan Digital Sky Survey (SDSS\footnote{\url{http://www.sdss.org/}}) on 2000 May 4, and were subtracted from the late-time $griz$ images. This procedure was needed in particular for the LCO images, because the seeing was typically poor and hence background contamination more severe.

We determined the colour terms for all filters using observations of standard fields. Specifically, we used standard photometric fields from \citet{1992AJ....104..340L}~for calibrating the Johnson-Cousins data, and fields covered by the SDSS DR 13 catalogue for the Sloan-filter observations. The photometric zero point of each image was calibrated with reference to a subset of stars in the field of the transient from the SDSS DR 13 catalogue. To determine the zeropoint of the Johnson-Cousins filter images, the Sloan-filter magnitudes of SDSS sources in the field were converted to Johnson-Cousins filters using the relations of \citet{2008AJ....135..264C}. The final optical magnitudes of AT~2017be are reported in Table~\ref{optical_bands}. We note that very few observations were obtained using the Johnson-Cousins $R$ and $I$ filters, hence these data were converted to Sloan-$r$ and $i$ SDSS systems following again the relations given in \citet{2008AJ....135..264C}.

\begin{table}
\caption{Parameters for the peak and the plateau.\label{maxdata}}
\begin{footnotesize}
\begin{tabular}{lccc}
\hline         
Filter& m$_\mathrm{peak}$ (mag)& m$_\mathrm{plateau}$~(mag) & $\Delta\mathrm{m}$ (mag) $^a$\\
\hline 
$B$ &  $18.46 \pm 0.07$ & $20.07 \pm 0.06$ &$1.61 \pm 0.09$\\
$V$ &  $18.05 \pm 0.05$ &  $19.03 \pm 0.05$ & $0.98 \pm 0.05$\\
$g$  &  $18.21 \pm 0.05$ &  $19.33 \pm 0.04$& $1.12 \pm 0.06$\\
$r $  &  $17.70 \pm 0.04$ & $18.63 \pm 0.04$& $0.93 \pm 0.04$\\
$i$  &$17.73 \pm 0.03$ & $18.38 \pm 0.03$& $0.65 \pm 0.03$\\
$z$ &  $17.45 \pm 0.35$ & $18.12 \pm 0.03$& $0.67 \pm 0.35$\\
\hline
\end{tabular}
\\[0.5ex]
$^a$ $\Delta\mathrm{m}$ is the magnitude difference between the averaged plateau and the peak.
\end{footnotesize}
\end{table}

\begin{table}
\caption{Decline rates in mag/100d for the lightcurves.  \label{gammaopt} }
\begin{footnotesize}
\begin{tabular}{lccc}
\hline   
 Filter &   $\gamma_1$ (mag/100d)&$\gamma_2$ (mag/100d) &$\gamma_3$ (mag/100d)\\  
  \hline
$B$ &$3.22 \pm 0.04$ & $ 0.61 \pm 0.06$& $ 2.80 \pm 0.07$ \\
$V$ & $1.90 \pm 0.03$ & $0.74  \pm 0.05$& $ 2.84 \pm 0.05$ \\ 
$g$ & $2.67 \pm 0.03$  &$0.40 \pm 0.04$ &$ 2.77 \pm 0.07$ \\
$r$ & $1.81 \pm 0.02$ &  $0.60 \pm 0.04$ & $ 2.29 \pm 0.03$ \\
$i$ & $1.18 \pm 0.02$  &  $0.89 \pm 0.03$ &$ 2.12 \pm 0.03$\\
$z$ &  $0.54 \pm 0.03$  & $0.61 \pm 0.03$ &$ 2.31 \pm 0.04$ \\
$J$ &       --         &$0.84 \pm 0.14$&$ 1.35 \pm 0.28$\\
$H$ &      --        &$0.66 \pm 0.14$& $ 1.64 \pm 0.27$\\
$K$ &     --         &$1.08\pm 0.15$ &$ 1.31 \pm 0.12$\\
\hline
\end{tabular}
\\[0.5ex]
 
\end{footnotesize}
\end{table}

NIR reductions include differential flat-field correction, sky subtraction and distortion correction. The reductions were based on a modified version of the external {\sc iraf} package NOTCam version 2.5\footnote{\url{http://www.not.iac.es/instruments/notcam/guide/observe.html\#reductions/}}. To improve the S/N, for each filter we combined multiple individual dithered exposures. NIR instrumental magnitudes of AT~2017be were measured with {\sc SNoOpy}, and then calibrated using sources in the field from the Two Micron All Sky Survey (2MASS\footnote{\url{http://irsa.ipac.caltech.edu/Missions/2mass.html/}}) catalogue. The resulting NIR magnitudes (in the Vegamag system) for AT~2017be are reported in Table~\ref{JHKcurves}.\\

\subsection{Lightcurves} \label{photoanalysis}
\begin{figure*} 
\centering
\includegraphics[width=2.0\columnwidth]{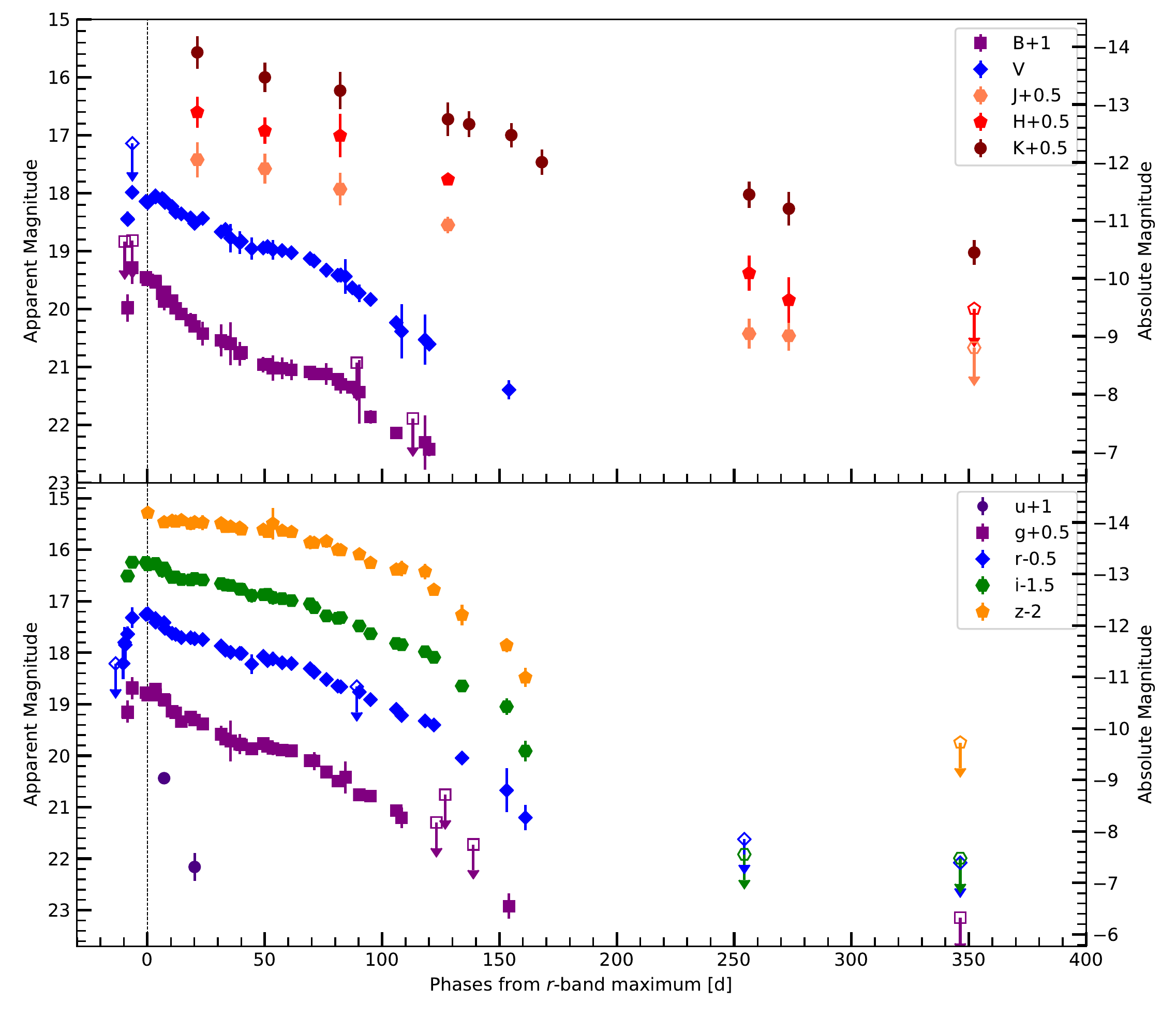}
  \caption{Multi-band lightcurves of the transient AT~2017be. Both the optical and NIR observations are shown in this Figure. Note that the $ugriz$ magnitudes are in the AB magnitude scale, while $BV$ and $NIR$ magnitudes are calibrated to the Vega magnitude scale. Some points are taken from published values (as reported in Table~\ref{optical_bands}). Detection limits at early and late phases are indicated with an empty symbol with a down-arrow. The errors for most data points are smaller than the marked symbols. }
\label{lightcurves}
\end{figure*}

The last pre-discovery non-detection reported by the KAIT is from 2016 December 29.479 UT (MJD = 57751.479), while the discovery epoch is 2017 January 6.508 UT (MJD = 57759.508). Therefore, it is reasonable to assume that the explosion of AT~2017be occurred between these two epochs. In the absence of more stringent indications, we adopt an explosion epoch of MJD = $57755 \pm 4$. Soon after the discovery of  AT 2017be, we initiated a photometric followup campaign which lasted for around one year. The resulting multi-band lightcurves of AT~2017be are shown in Figure~\ref{lightcurves}. In the forthcoming discussion, we will adopt the epoch of the $r$-band peak magnitude (MJD = 57769.8 $\pm$ 0.1) as the reference time for the light and colour curves (see Sect. \ref{coloranalysis}). This was obtained by fitting a 3rd-order polynomial to the early $r$-band lightcurve.  
        
The apparent magnitudes of AT~2017be in the different bands at maximum are listed in Table~\ref{maxdata}, while the decline rates in different phases of the evolution obtained through linear fits are reported in Table~\ref{gammaopt}. Similar to other ILRTs, AT~2017be rises rapidly and reaches the peak magnitude in all bands in less than two weeks. The blue-band lightcurves reach their luminosity peaks a few days earlier than the red-band ones (see Figure~\ref{lightcurves}). After maximum, the luminosity of AT~2017be decays quite rapidly in all bands until day $\sim$ 30, with the blue band (e.g., $BgV$) lightcurves usually fading more rapidly than the red bands (e.g., $ri$). Subsequently, the luminosity settles onto a plateau phase lasting approximately 30 days. The averaged plateau apparent magnitudes are also reported in Table~\ref{maxdata}. After the plateau phase, all lightcurves fade again more rapidly. From a morphological point of view, we can hence split the lightcurve evolution in three phases: from the peak to $\sim$ 30 days there is an initial fast decline ($\gamma 1$ $\approx$ 1-3 mag/100d, see Table~\ref{gammaopt}); from $\sim$ 30 days to $\sim$ 70 days a plateau ($\gamma 2$ $\approx$ 0.6-0.9 mag/100d), and from $\sim$ 70 days to the latest detections before solar conjunction a faster late decline ($\gamma 3 \approx$ 1.3-2.8 mag/100d). The transient was followed in photometry up to late phases, but only upper limits were obtained in all optical bands after solar conjunction (which lasted $\sim$ 100 days). Complementary NIR photometry is shown in Figure~\ref{lightcurves} and the data are reported in Table~\ref{gammaopt}. Unfortunately, we did not follow the transient in the NIR during the first month, hence the lightcurve rise and the subsequent NIR peak have not been monitored, while the initial decline, and the late decline phases have been observed. The NIR decline rates $\gamma 2$ (optical plateau-like period) and $\gamma 3$ are reported in Table~\ref{JHKcurves}. We included the earliest photometric points in our measurements of $\gamma 2$, and this may explain the relatively large decline rates found in the NIR lightcurves during the intermediate plateau phase (up to $\sim$ 80 days). We also note that the measured $K$-band decline rate at very late phases in the $K$ band is around 1.02 mag/100d, which is consistent with the theoretical $^{56}$Co decay of 0.98 mag/100d assuming complete gamma-ray and positron trapping. However, this is not a strong argument to support the $^{56}$Co decay powering mechanism at late times, as the spectral energy distribution (SED; see Section \ref{sed}) suggests the presence of a clear near-IR excess at these epochs probably due to emission from pre-existing dust in the CSM of AT~2017be. For this reason, we believe that the observed K-band decline match with the $^{56}$Co decay rate is likely a coincidence. \\


\subsection{Colour and Absolute Lightcurves} \label{coloranalysis}

\begin{figure} 
\includegraphics[width=1\columnwidth]{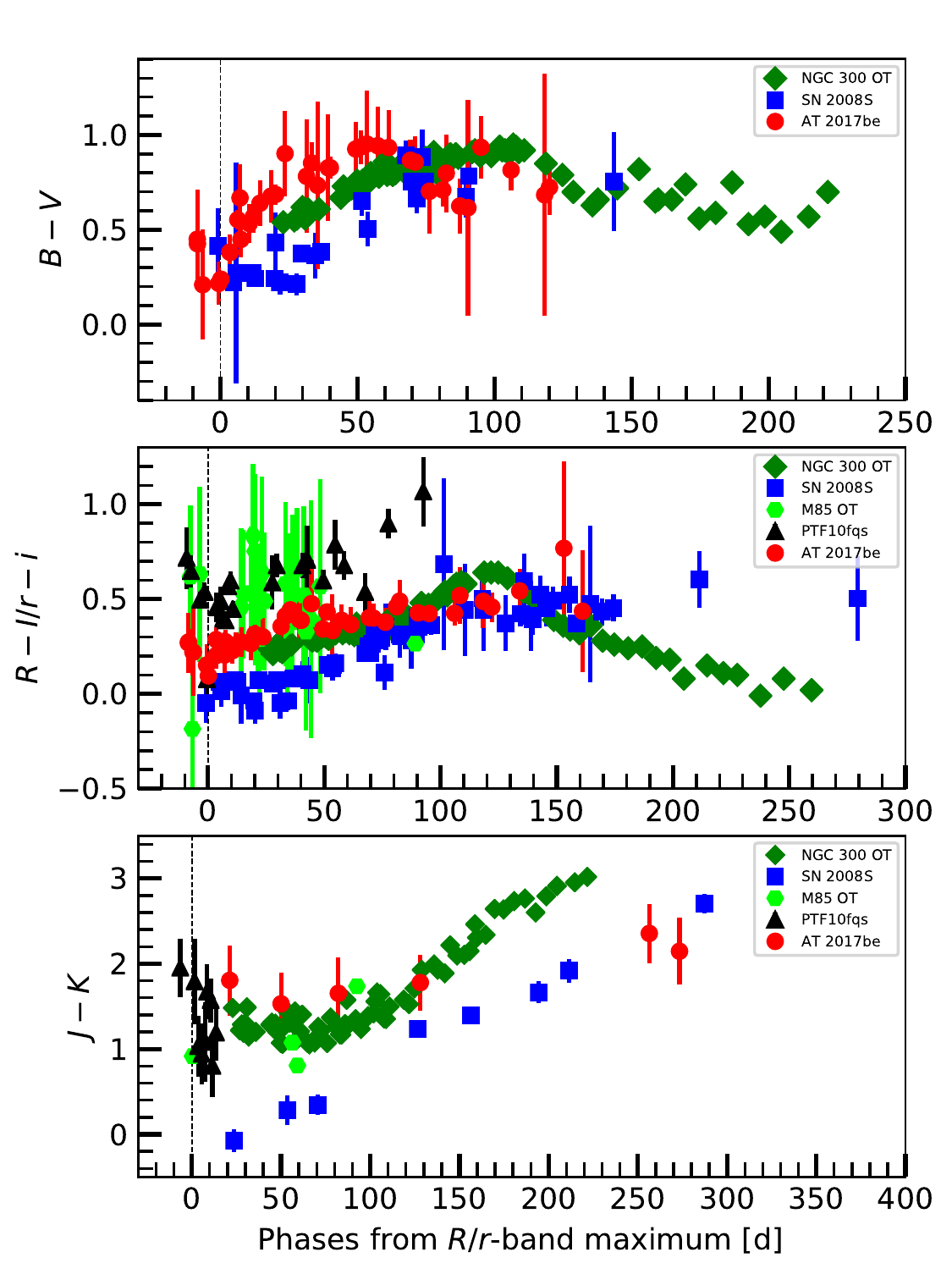}
\caption{Comparison of $B-V$, $R-I/r-i$ and $J-K$ colour curves of AT~2017be, SN~2008S, NGC~300~OT2008-1, M85~OT2006-1 and PTF10fqs. The data of SN~2008S are from \citet{2009MNRAS.398.1041B}, those of NGC~300~OT2008-1, M85~OT2006-1,  PTF10fqs are from \citet{2011ApJ...743..118H}, \citet[][]{2007Natur.447..458K} and \citet{2011ApJ...730..134K}, respectively. The adopted reddening values are given in the main text. The $r-i$ colours shown here are obtained from Sloan magnitudes, conveted to the Vega system. The errors on $J-K$ colour were not included in the NIR dataset for NGC~300~OT2008-1.} 
\label{colCurves}
\end{figure}

Figure~\ref{colCurves} shows the $B-V$ (upper panel), $R-I$/$r-i$ (middle panel), $J-K$ (lower panel) colour evolution of  AT~2017be along with that of other ILRTs, including SN~2008S \citep{2009MNRAS.398.1041B}, NGC 300~OT2008-1 \citep{2011ApJ...743..118H}, M85~OT2006-1 \citep[][]{2007Natur.447..458K} and PTF10fqs \citep{2011ApJ...730..134K}.

For SN~2008S,  we adopted a distance modulus  $\mu$ = 28.78 $\pm$ 0.08 $\rm{mag}$, a Galactic reddening \hbox{E($B-V$)$_\rmn{Gal}$ = 0.36}~mag  and a host galaxy reddening \mbox{E($B-V$)$_\rmn{host}$ = 0.32~mag} \citep{2009MNRAS.398.1041B}. For NGC 300~OT2008-1, we adopted a distance modulus of $\mu$ = 26.37 $\pm$ 0.14 $\rm{mag}$ \citep[e.g.,][averaging several values available in the literature with Cepheid estimations]{2010ApJ...715..277B, 2013AJ....146...86T, 2016AJ....151...88B}, a Galactic reddening \hbox{E($B-V$)$_\rmn{Gal}$ = 0.01}~mag \citep{2011ApJ...737..103S} and a host galaxy reddening \mbox{E($B-V$)$_\rmn{host}$ = 0.25~mag} (Na I D measurement), which agree with the estimates of \citet{2011ApJ...743..118H}. For M85~OT2006-1, \citet[][]{2007Natur.447..458K} estimated a distance modulus of $\mu$ = 31.33 $\pm$ 0.14 $\rm{mag}$ and a total reddening \mbox{E($B-V$)$_\rmn{total}$ = 0.14~mag}. Here we updated the distance modulus of $\mu$ = 31.00 $\pm$ 0.12 $\rm{mag}$ \citep{2013AJ....146...86T} and used the same reddening estimate. Finally, for PTF10fqs, \citet{2011ApJ...730..134K} adopted Galactic reddening \hbox{E($B-V$)$_\rmn{Gal}$ = 0.04}~mag only. Note that a non-negligible contribution from host galaxy reddening cannot be ruled out, although the host galaxy is an S0-type, as there is evidence of low level star formation \citep{2007Natur.449E...1P}. Here we adopt the \citet{2011ApJ...730..134K} extinction and update the distance modulus to $\mu$ = 30.77 $\pm$ 0.52 $\rm{mag}$ \citep{2009ApJS..182..474S}. 

At early phases, the transients included in our sample show a homogeneous colour evolution, which can be indicative of similar temperatures. The $B-V$ colour of AT~2017be evolves to progressively redder colours reaching 0.9 $\rm{mag}$ at about 60 days after  maximum, which suggests a significant decrease in the ejecta temperature. This is confirmed through the inspection of the spectral blackbody temperature (Section~\ref{spectroscopy}). Later on, the $B-V$ index moves back to a bluer colour and remains between $0.5\, \rm{mag}$ and $0.9\, \rm{mag}$ at later phases. This evolution of AT~2017be is similar to that of NGC 300~OT2008-1. Figure~\ref{colCurves} (middle) shows the $R-I$/$r-i$ colour evolution of the five ILRTs. The $r-i$ colour of AT~2017be has very little evolution to redder colours (reaching $0.3\, \rm{mag}$) until about day 30, while $B-V$ $\sim$ 0.7 mag at the same epoch (see the upper panel of Figure~\ref{colCurves}). The $r-i$ colour of AT~2017be experiences a monotonic rise from $0.1\, \rm{mag}$ to $0.5\, \rm{mag}$ during the entire follow-up campaign. We note that a significant dispersion is observed in the colour evolution of AT~2017be in the data presented by \citet[][see their figure 1]{2017arXiv171110501A}, while our data define a clearer colour evolution. However, the maximum colour of $\sim$ $0.5\, \rm{mag}$ in our $r-i$ colour curve is in agreement with their measurement. A comparison at later phases is more difficult due to the larger dispersion in the \citet{2017arXiv171110501A} data. The $J-K$ colour curves for our ILRT sample are shown in Figure~\ref{colCurves} (lower panel). We note that AT~2017be shows a NIR colour minimum of $1.5\, \rm{mag}$ at the phase of around 50 days, then it becomes progressively redder with time, reaching $2.2\, \rm{mag}$. The overall trend with $J-K$ colours becoming redder with time seems to be a common property for all ILRTs. 

\begin{figure} 
\includegraphics[width=1.0\columnwidth]{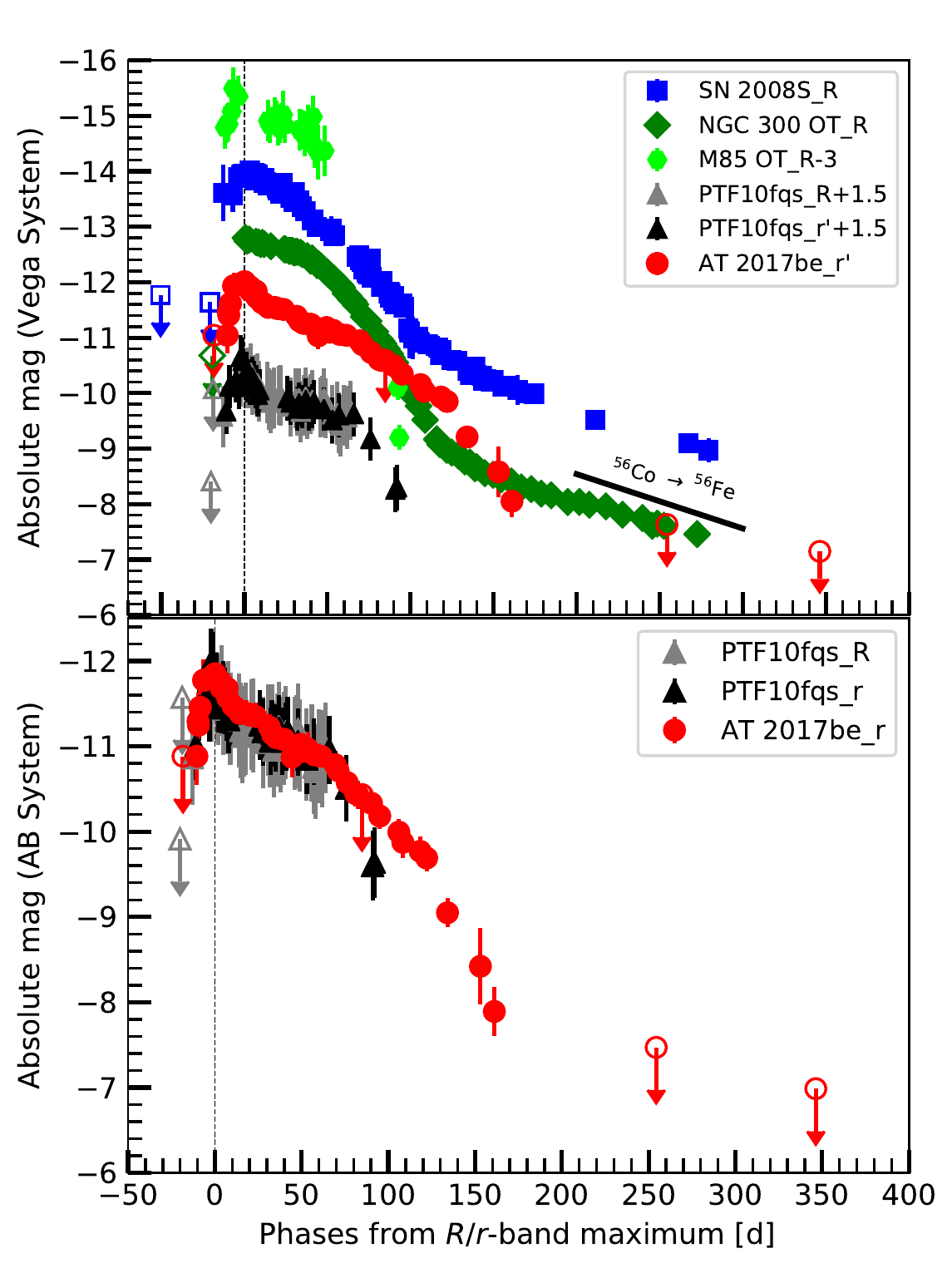}
\caption{$R$/$r$ band absolute lightcurves of AT~2017be and other ILRTs, including SN~2008S \citep{2009MNRAS.398.1041B}, NGC~300~OT2008-1 \citep{2011ApJ...743..118H}, M85~OT2006-1 \citep[][]{2007Natur.447..458K}  and PTF10fqs \citep{2011ApJ...730..134K}. In order to improve visualisation, arbitrary shifts are applied to M85~OT2006-1 and PTF10fqs as per the legend. We adopt the time of the $R$/$r$ band lightcurve peaks as our reference epoch. }
\label{absComparison}
\end{figure}

Figure~\ref{absComparison} shows the absolute $r$-band lightcurve of AT~2017be, compared with those of the ILRTs mentioned above. AT~2017be reached an absolute peak magnitude of $M_r$ = $-$11.98 $\pm$ 0.09 $\rm{mag}$ as measured using a 3rd-order polynomial fit to the lightcurve. A multi-band absolute lightcurve of AT~2017be was also shown in \citet[see their figure 15]{2017arXiv171110501A}. Their measurement of the Sloan $r$-band peak is around $-$11.4 \rm{mag}. This difference of over half a magnitude is mainly due to different adopted distances and (more marginally) to  the different extinction estimates. The \citeauthor{2017arXiv171110501A} apparent magnitude lightcurves are in good agreement with our data. The above value is intermediate between those of luminous novae and classical CC SNe (i.e., $-10 > M_{V} > -15$ mag), and is consistent with that expected for an ILOT \citep{2011MNRAS.415..773S, 2016ApJ...823L..23T}. The $r$-band lightcurve of AT~2017be has an asymmetric lightcurve peak, with a rapid rise but a modest post-peak decline, with the lightcurve settling onto a sort of plateau. AT~2017be is fainter than SN~2008S ($M_R$ = $-$14.24 $\pm$ 0.03 $\rm{mag}$) and NGC~300~OT2008-1 ($M_R$ = $-$13.23 $\pm$ 0.05 $\rm{mag}$), and nearly the same luminosity as M85~OT2006-1 ($M_R$ = $-$12.64 $\pm$ 0.50 $\rm{mag}$) and PTF10fqs (absolute peak magnitude $M_R$ = $-$11.56 $\pm$ 0.33 $\rm{mag}$; see the upper panel of Figure~\ref{absComparison}). Figure~\ref{absComparison} (lower panel) shows the $R$/$r$-band absolute magnitude evolution of  AT~2017be and PTF10fqs in the ABmag scale \citep{1983ApJ...266..713O}. The comparison shows the striking similarity in the early lightcurve evolution for these two ILRTs, which reach approximately the same maximum magnitudes.

\begin{figure} 
\includegraphics[width=1.0\columnwidth]{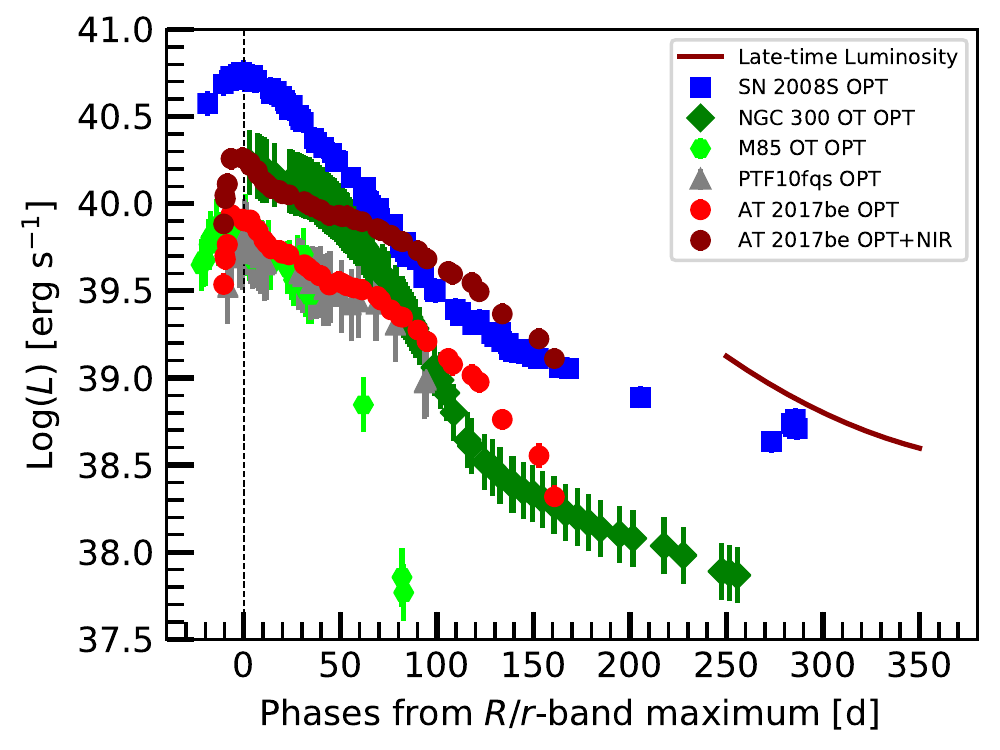}
\caption{Quasi-bolometric lightcurves of AT~2017be, SN~2008S \citep{2009MNRAS.398.1041B}, NGC~300~OT2008-1 \citep{2011ApJ...743..118H}, M85~OT2006-1 \citep[][]{2007Natur.447..458K}and PTF10fqs \citep{2011ApJ...730..134K}. For the purpose of comparison, only the Johnson/Cousins $BVRI$and SDSS $gri$ filters  were used. Note that for AT~2017be we also computed the $u$-to-$K$ band lightcurve (shown with dark red dots). Errors shown account for the uncertainties on photometric measurements, distance and reddening estimates, and SED fitting. }
\label{Bolo}
\end{figure}

\subsection{Quasi-bolometric Lightcurve} \label{boloanalysis} 

Quasi-bolometric luminosities are obtained by converting the available magnitudes to flux densities, then correcting these fluxes with the adopted extinction coefficients, and finally integrating the SED over the entire wavelength range. We assume that the flux at the integration limits (at effective wavelengths shorter than the $u$ band and longer than the $K$ band ) is zero. At some epochs, in a given filter, in which the photometric data are not available, we estimate the flux contribution of the missing bands by interpolating between epochs with available data or through an extrapolation from the earlier or later available epochs, assuming a constant colour evolution. We calculated the pseudo-bolometric lightcurve of AT~2017be and other ILRTs using only the $BVRIgri$ bands. This allows for meaningful comparisons among ILOTs, for which often $u, z$~or $NIR$ bands were not available. In the case of AT~2017be we also computed a more complete quasi-bolometric lightcurve including the contribution to the total luminosity of the $u$ and the NIR ($z, J, H, K$) bands. The resulting quasi-bolometric lightcurves are shown in Figure~\ref{Bolo}. 

The quasi-bolometric lightcurve peak of AT~2017be is $\sim$ $8 \times 10^{39}$ erg s$^{-1}$, including the optical bands only; this value increases to $\sim$ 1.8 $ \times10^{40}$ erg s$^{-1}$ including also the $u$ and NIR bands contributions ($\sim$ 55.6\%). The peak luminosities of SN~2008S, NGC~300~OT2008-1, M85~OT2006-1, and PTF10fqs (considering optical bands only) are  about $6 \times10^{40}$ erg s$^{-1}$, $1.8 \times 10^{40}$ erg s$^{-1}$,  $10^{40}$ erg s$^{-1}$ and $6 \times 10^{39}$ erg s$^{-1}$ respectively. The lightcurve shape of AT~2017be is different from those of SN~2008S and NGC~300~OT2008-1, having a somewhat narrower peak and a post-peak plateau, and is very similar to those of M85~OT2006-1 and PTF10fqs, in overall shape, peak luminosity and post-peak lightcurve flattening. The main difference is in the duration of the plateau. While the post-plateau decline in AT~2017be begins about 70 days after maximum light, and at a similar phase in PTF10fqs, for M85~OT2006-1 the plateau only lasts for around 30--35 days. The post-plateau decline of AT~2017be is very rapid, with a rate that is faster than the decay of $^{56}$Co (see Fig,~\ref{Bolo}). Unfortunately, we could not follow the object to observe an optical flattening onto the radioactive tail, because the transient disappeared behind the sun. However, we observed again the field a few months later, when it became visible again before the sunrise, and noticed that AT~2017be was no longer visible in the optical bands, while it was clearly detected in the NIR bands. 

As we have a  limited lightcurve information for  AT~2017be at late phases, we cannot securely estimate its bolometric light curve, which is essential to constrain the ejected $^{56}$Ni mass . 
In addition, AT~2017be has no mid-IR monitoring, and we know that  similar objects like SN~2008S and NGC~300~OT2008-1 have a non-negligible mid-IR contribution at late phases, with the peak of their SEDs shifting from the optical to the mid-IR  \citep{2011ApJ...741...37K}. Given that SN~2008S and NGC~300~OT2008-1 have relatively complete datasets spanning from the optical to mid-IR and extended to late phases, we use their SEDs to estimate the relative flux contribution of the different wavelength regions.
Assuming that the SED evolution of AT~2017be is similar to that of the two template objects, we can in this way constrain the contribution of the missing bands and estimate a reliable bolometric light curve also for the former object.

We first used SN 2008S as template for AT 2017be, in view of the similar $J-K$ colour evolution. We estimated the NIR luminosity of SN 2008S from its light curve, and adopted
 its model bolometric lightcurve obtained by \citet{2011ApJ...741...37K} through the relation: L$_{bol}$ = $L_0$ $\exp \left(-t/t_0\right)$ + $L_1$, with $L_0 \simeq 10^{7.3} L_\odot$, $L_1 \simeq 10^{5.8}L_\odot$ and $t_0\simeq 48$~days. Hence we computed the L$_{NIR}$/L$_{bol}$ ratio for SN~2008S to be $\approx$ 0.31 - 0.36, from  $\sim$ 245 to 345 days after maximum. Finally, we obtained the bolometric light curve of AT~2017be using its  L$_{NIR}$ values obtained from the light curve at late phases assuming the same L$_{NIR}$/L$_{bol}$ ratio as SN~2008S. The resulting bolometric lightcurve for AT 2017be at late phases is also shown in Figure~\ref{Bolo} with a solid line.  
 
Assuming that the energy from $\gamma$ rays and positrons is fully trapped and thermalised, we constrain the $^{56}$Ni mass deposited by AT~2017be by comparing its late-time bolometric luminosity with that of SN 1987A at the same epochs ($\sim$ 260-360 days after their explosion). The luminosity of SN 1987A was calculated by \citet{1988MNRAS.234P...5W}, including the U- to M-band contribution (we will assume that the contribution to the total luminosity from the regions outside this range is negligible). The $^{56}$Ni mass systhesized by AT~2017be ($\sim$ 7 $\times$ $10^{-4}$\msun) is obtained through the relation:  
\begin{equation}
\mathrm{M}\left(^{56}\mathrm{Ni}\right)_\mathrm{AT~2017be}~$=$~0.073\mathrm{M_{\odot}}\times \left(L_\mathrm{AT~2017be}\left(t\right)\over L_\mathrm{SN~1987A}\left(t\right)\right).
\end{equation}
As a further test, we can use the SED of NGC~300~OT2008-1 as a template, adopting the same algorithm as SN 2008S, and we find M($^{56}$Ni) $\sim 9 \times 10^{-4}$\msun\ . Averaging the above estimates, we infer M($^{56}$Ni) =  8 ($\pm$ 1) $\times 10^{-4}$\msun\ for AT 2017be. However, we remark that additional contributions to the late bolometric lightcurve from interaction between ejecta and CSM, or even IR echoes cannot be definitely ruled out (see the further discussion in Section \ref{sed}). In that case, the above  M($^{56}$Ni) value should be regarded as an upper limit.

Integrating the bolometric lightcurve over the entire observational campaign ($\sim$ 370 days) of AT 2017be, we can estimate a reliable value for the total radiated energy. We have a good early time coverage in the optical and NIR bands
at early phases. Hence, we can estimate the contribution of the missing bands using again both  SN~2008S and NGC~300~OT2008-1 templates. In the two cases, we obtain $\sim$ $1.19 \times  10^{47}$ erg and $\sim$ $1.31 \times 10^{47}$ erg as energy radiated by AT~2017be, with an average of 1.25 ($\pm$ 0.06) $\times 10^{47}$ erg.

Although ECSNe are expected to arise from stars in a very narrow range of masses, and the mass of $^{56}$Ni only marginally affect the observed luminosity (in particular, at early phases), the quasi-bolometric lightcurves of ILRTs are expected to be reasonably similar \citep[e.g.,][]{2016MNRAS.461.2155M, 2009ApJ...705L.138P, 2006A&A...450..345K}. Indeed, as shown in Figure~\ref{Bolo}, the lightcurves of all our ILRTs are indeed broadly similar. The differences between these events can potentially be explained by a number of factors, including the mass-loss history (hence the geometric distribution and density profile of the CSM), the H recombination timescale, and the influence of $^{56}$Ni distribution inside the ejecta.

\subsection{Spectral energy distribution analysis} \label{sed} 
\begin{figure} 
\includegraphics[width=1.0\columnwidth]{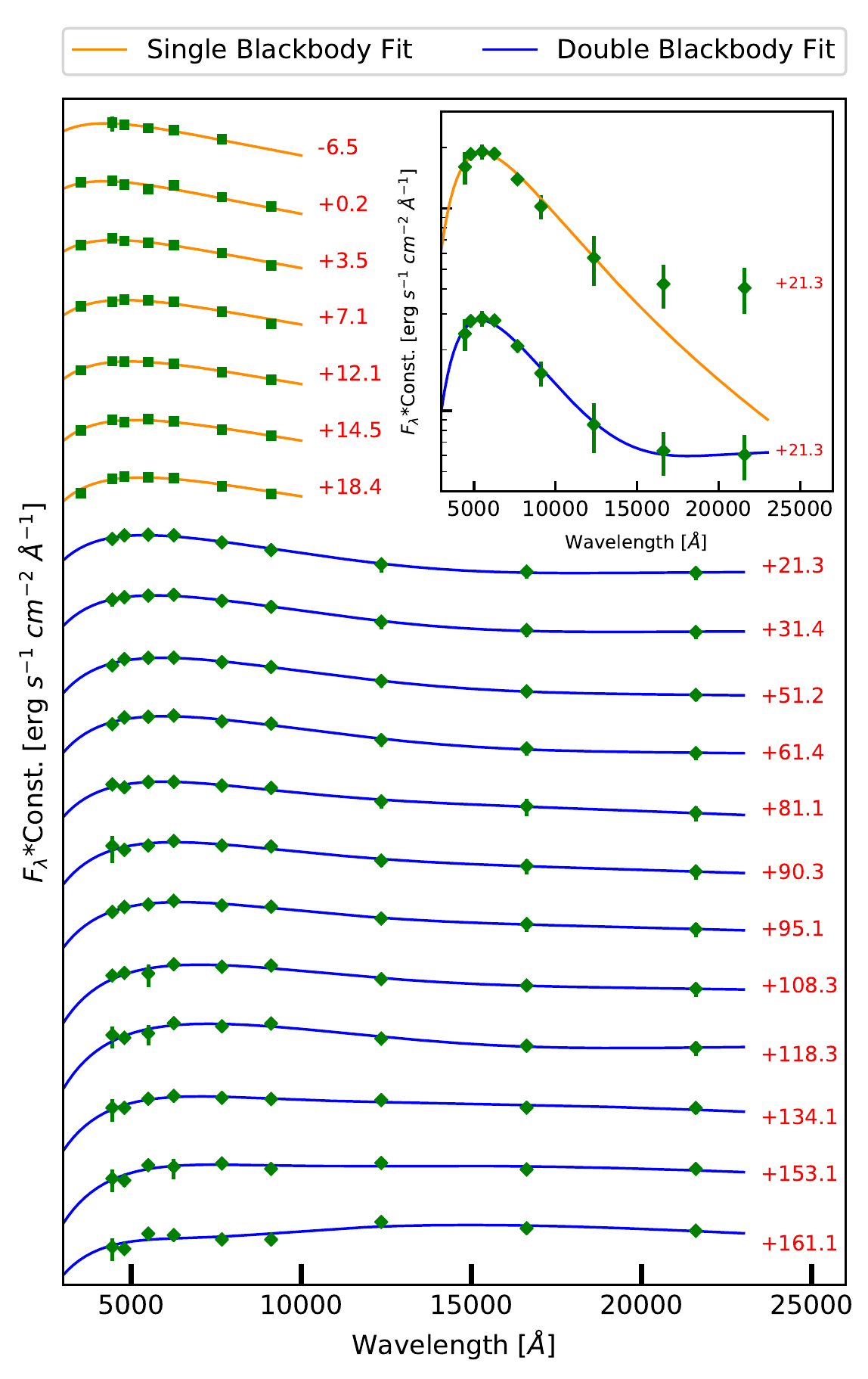}
\caption{The observed SED evolution of AT~2017be. The fluxes have been corrected for reddening. We adopt the value of MJD = 57769.8 $\pm$ 0.1 (Sloan $r$-band maximum) as our reference epoch. }
\label{sedevolution}
\end{figure}

\begin{figure} 
\includegraphics[width=1.0\columnwidth]{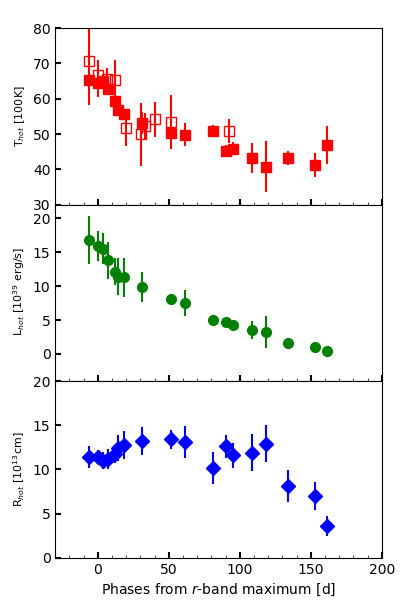}
\caption{Temporal evolution of parameters determined from the SED of AT~2017be. In the upper panel, empty squares indicate the temperature evolution deduced from the blackbody fit to the spectral continuum. The uncertainty in the inferred blackbody temperature has been estimated based on the uncertainties in flux in each bandpass.  We adopt MJD = 57769.8 $\pm$ 0.1 (Sloan $r$-band maximum) as our reference epoch. }
\label{Pevolution}
\end{figure}

The temporal evolution of the SED of AT~2017be is shown in Figure~\ref{sedevolution}. We did not collect NIR observations until about three weeks after maximum, so our SED at early epochs is based solely on optical data. These early epochs are well represented by a single blackbody, which is fit to the optical data, whose temperature declines rapidly from $\sim$ 6500~K to $\sim$ 5500~K by $\sim$+20~days (see the upper panel of Figure~\ref{Pevolution}). The first optical+NIR SED is obtained at  $+21$~days from maximum, which reveals a notable NIR excess over a single black body (see inset in the upper-right corner of Figure~\ref{sedevolution}). To investigate the nature of the NIR excess, we carried out a simultaneous two-component fit to the optical+NIR SEDs consisting of two (hot plus warm components) black-body functions. Occasionally, for some epochs fluxes for unobserved bands were estimated through an interpolation procedure. At these later epochs (after +20 days), the temperature of the hot component slowly declines from about 5500~K to 4300~K at about 160 days. As expected, the above temperature estimates are in agreement with those obtained from the blackbody fit to the continuum of the spectra obtained at the similar epochs (see Section \ref{evolution_temperature}, and Figure \ref{Pevolution}, upper panel). Furthermore, the derived temperature is consistent with the hydrogen recombination temperature. The temperature of the warm blackbody is apparently constant over the 3 months (from $\sim$ 20 days to 150 days) of monitoring, within substantial errors, T = 1200 $\pm$ 200 K, indicative that this component is very likely due to thermal emission from surrounding dust. 

From luminosity and temperature we derive the radius of AT 2017be using the Stefan-Boltzmann law. The evolution of the wavelength-integrated luminosity of the hot component is shown in the middle panel of Figure~\ref{Pevolution}. It decreases from $\sim$ 1.7$ \times10^{40}$ erg s$^{-1}$ at $-6.5$ days to $\sim$ 4.5$ \times10^{38}$ erg s$^{-1}$ at about three months from maximum. The radius of the hot component (see the lower panel of Figure~\ref{Pevolution}) slowly increases from 1.1 $\times$ 10$^{14}$ cm near maximum light to 1.3 $\times$ 10$^{14}$  cm at about one month after peak luminosity (see lower panel in Figure~\ref{Pevolution}). After this until about four months past-maximum the radius decreases from about 1.4 $\times$ 10$^{14}$  cm to 1.2 $\times$ 10$^{14}$ cm, before dropping more rapidly to about 4 $\times$ 10$^{13}$cm at 161 days. The time evolution of the luminosity as well as the blackbody radius and hot temperature are reported in Table~\ref{sedparamters}, and shown in Figure~\ref{Pevolution}.  

NIR excess emission has been observed in several SNe and is often attributed to the thermal emission from warm dust heated by the SN. On the other hand, \citet{2011ApJ...741...37K} investigated the evolving SEDs of SN 2008S and NGC~300~OT2008-1, extending to mid-IR wavelengths, and found that the transient SEDs are largely determined by the dust reformation beyond the shock front.

Assuming the near-infrared emission is indeed due to dust, we fit a modified black-body function \citep{1983QJRAS..24..267H} to the optical+NIR SEDs. 
We fit for the dust temperature, $T_{\mathrm{d}}$, and dust mass, $M_{\mathrm{d}}$ adopting the formalism as described in \citep{2017ApJ...849L..19G}, which assumes that the dust mass absorption coefficient, $\kappa_{\mathrm{abs}}(\nu, a)$ (in units of [cm$^{2}$ g$^{-1}$]) behaves as a $\lambda^{-x}$ power law in the wavelength range 0.9--2.5 $\mu$m, with $x$ as the power-law slope. Here we perform fits for either $x$ = 1.2 or 1.5 and adopt $\kappa_{\mathrm{abs}}$($\lambda$ = 1 $\upmu\mathrm{m}$) = 1.0 $\times$ 10$^{4}$ cm$^{2}$ g$^{-1}$ (appropriate for carbonaceous dust, e.g. \cite{1991ApJ...377..526R}). 

The results for the temperature and radius of the black body function for the supernova are identical to the results obtained from the pure hot and warm black body fits. However, here we obtain a slightly lower temperature ($\sim$ 1000 K) for the dust component which also shows no evolution over time. The amount of dust decreases over time from a few times 10$^{-5}$ \msun~to a few times 10$^{-6}$ \msun~in the case of carbonaceous dust. However, the low dust temperatures do not rule out other dust species such as silicates. In the case of silicate dust the estimate of the dust mass would be about an order of magnitude larger.

Unfortunately, in the absence  of longer wavelength observations to better constrain the dust component we can only derive an upper limit to the dust temperature and a lower limit on the dust mass. From the analysis of the SED of AT 2017be SED, we infer that pre-exisisting dust from the progenitor likely survived in the circumstellar environment, and was probably heated by the SN. This dust has to be distributed far from the progenitor and such that it does not contribute to additional extinction along the line of sight towards the SN (see the extinction discussion in Section \ref{host}). This implies that either the dust mass is only slightly above our lower limit, or that the dust distribution of AT 2017be may be similar to that of V838 Mon \citep[e.g.,][]{2004A&A...414..223T, 2012A&A...548A..23T}, in form of clumpy, complex structure. \\

\begin{table*}
\caption{Properties of the hot blackbody component fit to the uBVgrizJHK bands of AT~2017be. The warm component 
does not show a significant evolution within its uncertainties. Uncertainties are given in parentheses.}
\label{sedparamters}
\begin{tabular}{@{}cccccc@{}}
\hline
Date & MJD & Phase$^a$  &   Luminosity (hot)                                     &  Temperature (hot)    &    Radius (hot)         \\ 
        &          & (d)                &  (10$^{39}$erg~s$^{-1}$) &  (K)               &    (10$^{13}$~cm)    \\ 
\hline
20170110$^b$ & 57763.3 & $-6.5$   &  16.8 (4.0)     &  6530 (190)   &   11.4 (1.0)        \\
20170116$^b$ & 57770.0 & +0.2      &   15.9 (2.0)    & 6430 (390)    &   11.4 (1.0)    \\
20170120$^b$ & 57773.3 & +3.5     &    15.5 (2.0)   & 6510 (170)   &  11.0  (1.0)      \\
20170123$^b$ & 57777.0 & +7.1     &    13.8 (3.0)  & 6280 (170)  &  11.2 (1.0)       \\    
20170128$^b$ & 57781.9 & +12.1   &    12.1  (2.0)    &  5930  (70) &   11.7 (1.0)   \\
20170131$^b$ & 57784.3 & +14.5   &    11.4  (3.0)     & 5680  (130) &  12.4 (2.0)   \\
20170204$^b$ & 57788.3 & +18.4   &    11.3  (3.0)  &  5580 (140)  & 12.8 (2.0)    \\  
20170207 & 57791.1& +21.3    &    10.9  (3.0)    & 5540 (220)    &  12.8 (1.0)   \\
20170217 & 57801.2& +31.4    &    9.9  (2.0)    & 5310 (300)    &  13.2 (2.0)   \\
20170309 & 57821.0 & +51.2   &    8.1  (1.0)      &  5020 (100)   &  13.4 (1.0)    \\
20170319 & 57831.2 & +61.4   &    7.6  (2.0)     &  4990 (320)    &   13.1 (2.0)       \\
20170407 & 57850.9 & +81.1   &    5.0  (1.0)      & 5100 (170)  &  10.2 (2.0)   \\
20170417 &57860.1 & +90.3    &    4.8   (1.0)     & 4530 (160)   &  12.6 (1.0) \\
20170421 &57864.9  & +95.1   &   4.3   (1.0)     & 4590 (200)    &  11.6 (1.0)  \\
20170505 &57878.1  & +108.3 &   3.5   (1.0)     & 4320 (420)     &  11.9 (2.0)   \\
20170515 &57888.1  & +118.3 &   3.3   (2.0)     & 4070 (720)   &  12.9 (2.0)  \\
20170530 &57903.9  & +134.1 &   1.6  (0.3)     & 4310 (200)    &  8.1 (2.0)   \\
20170618 &57922.9  & +153.1 &   1.0  (0.3)     & 4130 (330)    &  7.0 (2.0)  \\
20170626 &57930.9  & +161.1 &   0.4   (0.2)     & 4690 (540)    &  3.6 (1.0)  \\
\hline
\end{tabular}
\medskip

$^a$ Phases are relative to $r$-band maximum lightcurve, on MJD = 57769.8 $\pm$ 0.1.\\ 
$^b$ Data are obtained by fitting a single blackbody function, as we have no contemporaneous NIR observations.\\
\end{table*}

\section{Spectroscopy} \label{spectroscopy}
Our spectroscopic follow-up campaign started on 2017 January 10, and lasted until 2017 April 18, hence covering almost 100~days. Spectroscopic observations were performed using the 2-m Faulkes North Telescope of LCO with FLOYDS, the 2.56-m NOT with ALFOSC and the 10.4-m Gran Telescopio Canarias (GTC) with OSIRIS. Details on the instrumental configurations and  information on our spectra of AT~2017be are reported in Table~\ref{speclog}. 

\begin{table*}
\caption{Log of the spectroscopic observations of AT~2017be.}
\label{speclog}
\begin{tabular}{@{}cccccccc@{}}
\hline
Date &   & Phase$^a$  & Telescope+Instrument & Grism  & Spectral range & Resolution & Exp. time   \\ 
        &          & (d) &                               &                            & ($\ang$)                  & ($\ang$)           & (s)                \\ 
\hline
20170110 & 57763.4 & $-6.4$ &    LCO+FLOYDS      & Channels red+blue & 3600-10000 & 15.5 & 3600  \\
20170117 & 57770.0 & +0.2 &    NOT+ALFOSC                 & gm4                             &  3400-9700 &  15  & 3600 \\
20170123 & 57776.2 & +6.4 &    NOT+ALFOSC                 & gm4                          & 3390-9650  &  18  &  3512 \\
20170128 & 57781.9 & +12.1 &    NOT+ALFOSC                & gm4                          & 3500-9590  &  15  &  3600\\
20170205 & 57790.0 & +20.1 &    NOT+ALFOSC                & gm4                          & 3400-9650  &  18  &  3600\\
20170216 & 57800.0 & +30.2 &    GTC+OSIRIS                       & R2500R                    &  5580-7680 &  3.5    &  1800 \\
20170219 & 57803.0 & +33.2 &    NOT+ALFOSC                & gm4                          & 3400-9700  &  14  &  3600\\
20170225 & 57809.9 & +40.1 &    NOT+ALFOSC                & gm4                          & 3700-9650  &  19  &  3600\\
20170308 & 57821.0 & +51.2 &    NOT+ALFOSC                & gm4                          & 3400-9650  &  18  &  3600\\
20170418 & 57861.9 & +92.1 &    GTC+OSIRIS                       & R1000R                    &  5100-10400 &  8    &  2$\times$900 \\

\hline
\end{tabular}

\medskip

$^a$ Phases are relative to $r$-band maximum light, on MJD = 57769.8 $\pm$ 0.1.\\ 
\end{table*}

\begin{figure*} 
\centering
\includegraphics[width=2\columnwidth]{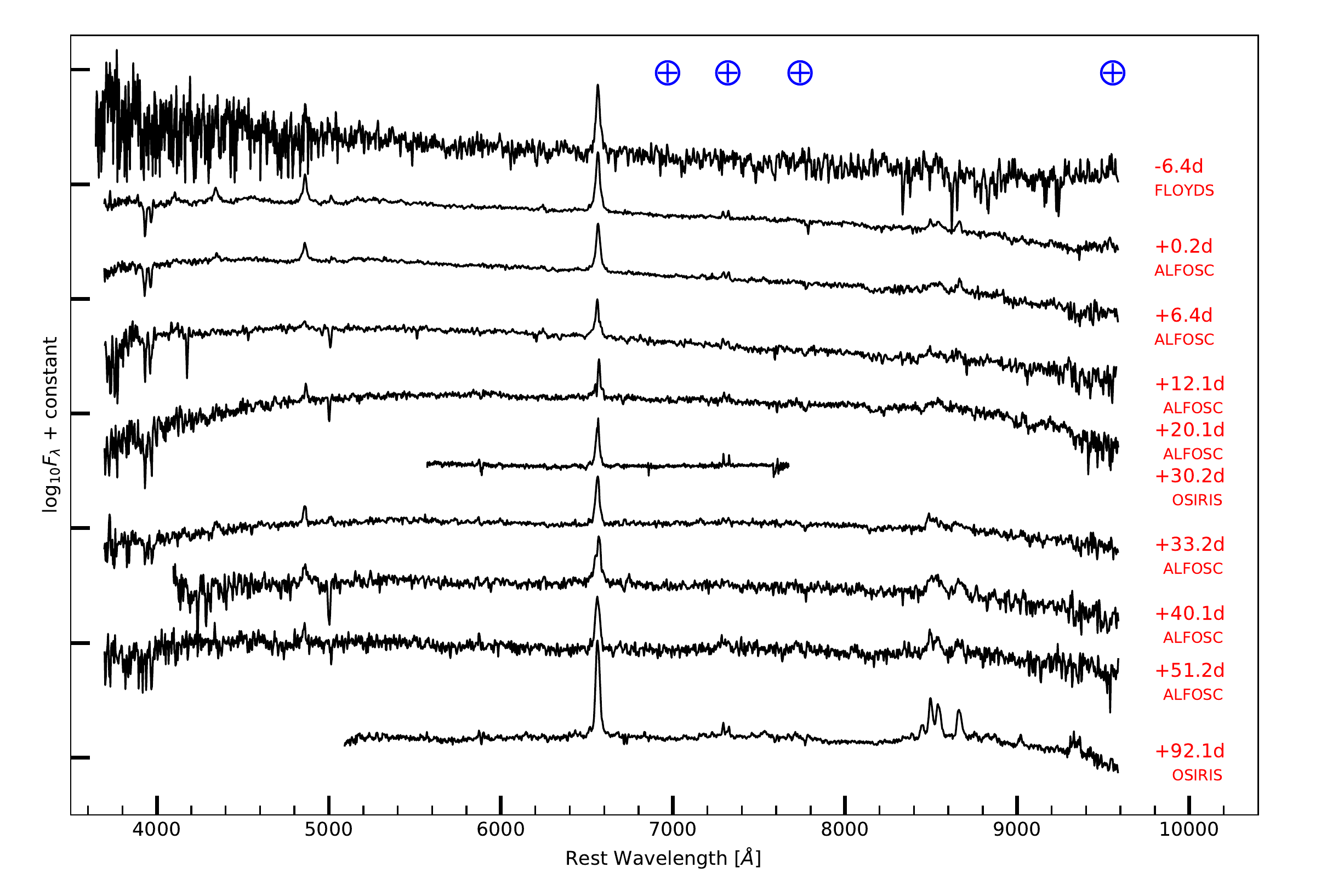}
\caption{Spectroscopic evolution of AT~2017be from 2017 January 10 (MJD = 57763.38) to  April 18 (MJD = 57861.93). The epochs reported to the right of each spectrum are relative to the Sloan r band maximum. The positions of the strongest telluric absorption bands are marked with the $\bigoplus$ symbol; all spectra have been redshift-corrected. }
\label{all_spectra}
\end{figure*}

The spectra of AT~2017be were reduced using standard tasks in {\sc iraf}\footnote{NOT+ALFOSC spectra were reduced using a dedicated graphical user interface (ALFOSC GUI) developed by E. Cappellaro (http://sngroup.oapd.inaf.it/foscgui.html).}. Overscan, bias, and flat-field corrections were applied using a similar procedure as for the photometric images. From the two-dimensional frames, we performed an optimal extraction of one-dimensional spectra using the package APALL. The one-dimensional spectrum was wavelength-calibrated using spectra of comparison lamps obtained during the same night and with the same instrumental setup. The accuracy of the  wavelength calibration was checked by measuring the observed wavelengths of a few night-sky lines, and (if necessary) a constant shift was applied to match the expected line wavelengths. Spectra of selected spectro-photometric standard stars were obtained during the same night as the SN observation to allow flux calibration of the spectra of AT~2017be. The precision of the flux calibration was verified using the available coeval photometric data. When some flux discrepancy was found, a scaling factor was applied to the spectrum to obtain a good match with our broadband photometry. Finally, the strongest telluric absorption bands of O$_2$ and H$_2$O in the spectra of AT~2017be were removed using the spectra of early-type standard stars.

In order to properly analyse our spectra of AT~2017be, we applied a  correction to account for the line-of-sight reddening contribution (see Section \ref{host}), and a Doppler shift to place the spectra in the host galaxy rest frame. The spectral resolutions, reported in Table~\ref{speclog}, were obtained by measuring the average widths of night sky lines.

\subsection{Spectral evolution}\label{evolution_temperature}
The spectra of AT~2017be show little evolution over the entire monitoring period (from 7 days to almost 100~days after the explosion, see Figure~\ref{all_spectra}).  All spectra consist of a nearly featureless continuum with strong Balmer emission lines. The [Ca {\sc ii}] doublet at 7291,7324~\AA~and the Ca~{\sc ii} NIR triplet are also detected. The P-Cygni absorption troughs, which usually characterise SN explosions or the presence of CSM material if narrow, are observed in our highest resolution GTC (R2500R) spectra. The H$\alpha$ line possibly shows a multicomponent profile, although in most  spectra of AT~2017be the line is well reproduced by fitting a single Lorentzian  profile (see Sect. \ref{evolution_H}).   

The spectra are characterised by a significant contribution of the continuum to the total flux, hence we estimated the temperature of the photosphere through a black-body fit. The black-body temperatures from spectra and imaging are consistent within the errors (see the upper panel of Figure~\ref{Pevolution}). We note that this temperature evolution is consistent with that of hydrogen recombination in normal Type II CC SNe \citep[e.g.,][]{1997ARA&A..35..309F}. The spectral blackbody temperature values are reported in Table~\ref{spectra_para} and are plotted in Figure~\ref{Pevolution} (upper panel).

\subsection{Line identification} \label{evolution_identification}

\begin{figure*} 
\centering
\includegraphics[width=2\columnwidth]{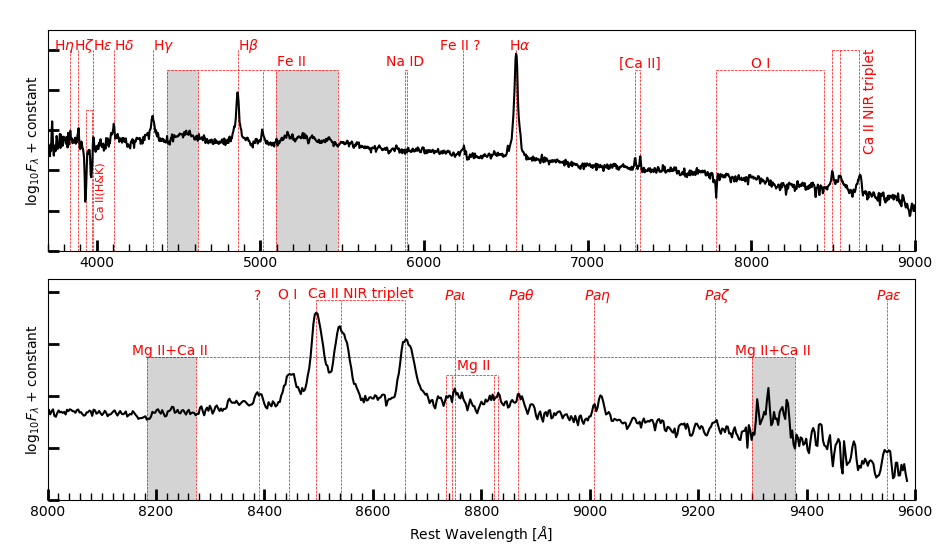}
\caption{Upper panel: Identification of the  strongest emission lines in the early-time ($\thicksim$1 day) NOT+ALFOSC spectrum of AT~2017be. Lower panel: Identification of the most prominent emission lines in the late-time ($\thicksim$90 days) GTC+OSIRIS spectrum of AT~2017be. Both the spectra have been redshift-corrected.}
\label{all_spectra_ident}
\end{figure*}

The line identification was performed on two high S/N spectra: an early epoch (+0.2 days) ALFOSC spectrum, and a late-time (92.1 days) GTC spectrum. The identification is shown in Figure~\ref{all_spectra_ident}. H$\alpha$ and other Balmer series emission lines are clearly present. Paschen series lines are also visible at the reddest wavelengths in our highest S/N spectra. Additionally, weak lines of Fe~{\sc ii}, Na~{\sc i}~D, O~{\sc i} , Ca~{\sc ii} H$\&$K and the Ca {\sc ii} NIR triplet are identified at their rest wavelengths. A possible Fe~{\sc ii} line is visible from first spectrum up to three weeks after maximum light (see the question mark in the upper panel of Figure~\ref{all_spectra_ident}). 

\begin{figure*} 
\centering
\includegraphics[width=2\columnwidth]{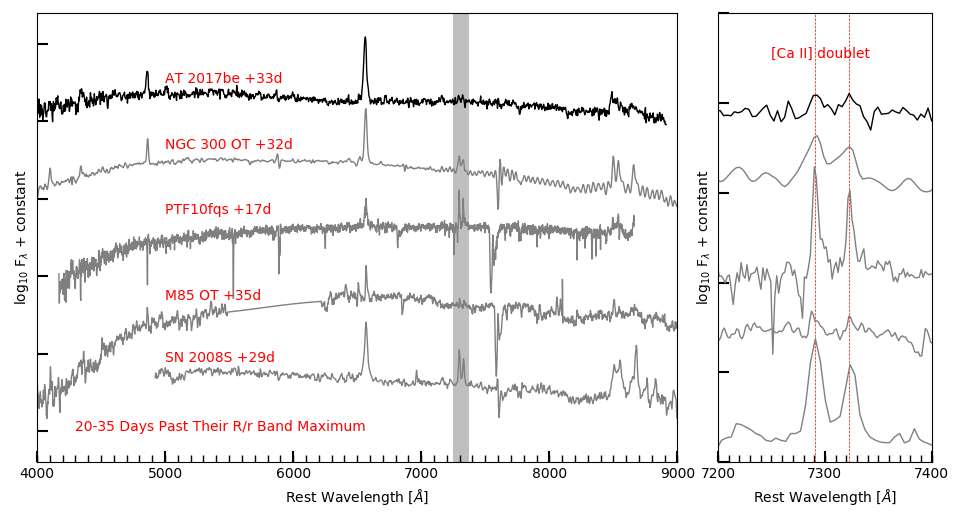}
\caption{Comparison of spectra of those of ILRTs obtained around the same epoch after their $R/r$ band maximum. All the spectra are taken from  Cai et al. in prep, \citet{2011ApJ...730..134K}, \citet[][]{2007Natur.447..458K} and \citet{2009MNRAS.398.1041B}. The [Ca~{\sc ii}] doublet lines are marked in the grey shaded area in the left panel. The right panel shows the detailed profiles. The spectra are all redshift-corrected and extinction-corrected.}
\label{comparison}
\end{figure*}

We also identify prominent [Ca~{\sc ii}] doublet lines at $\lambda\lambda$7291,7324, which are visible in all spectra of AT~2017be (see Figure~\ref{all_spectra}), and which are seen in all the ILRTs. In Figure \ref{comparison}, we compare spectra of AT~2017be and other four ILRTs at similar phases (marked with red labels in the figure). The striking spectral match of AT~2017be with the SN~2008S-like family is evident not only from the detection of the [Ca {\sc ii}] doublet (see right panel in Figure \ref{comparison}), but also from the presence of other shared spectral features (both in absorption and emission), providing further observational evidence to support our classification of AT~2017be as an ILRT. 

\begin{figure} 
\includegraphics[width=1.0\columnwidth]{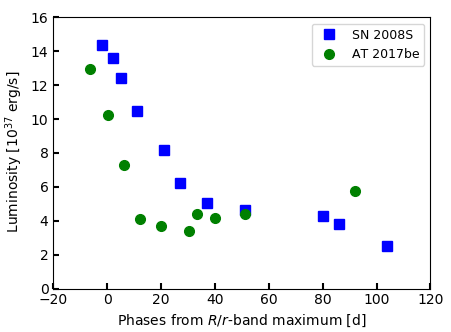}
\caption{Comparison between H$\alpha$ luminosity evolution in AT~2017be and SN~2008S. Values for SN~2008S are taken from the literature \citep[][see their Table 10]{2009MNRAS.398.1041B}. We conservatively estimate the uncertainties to be around 20 percent.}
\label{FWHM}
\end{figure}

\subsection{Evolution of the line profiles} \label{evolution_H}

In order to search for signatures of ejecta-CSM interaction we analysed the evolution of the line profiles, along with the velocity and the luminosity of H$\alpha$ at different phases. Multiple line components in spectra are normally associated with interaction between supernova ejecta and circumstellar wind, with the different components resulting from multiple emitting gas regions \citep{1993MNRAS.262..128T}. The classical interpretation is that the narrow component of H$\alpha$ is emission from unshocked CSM, while broader line components are signatures of shocked material and/or freely-expanding SN ejecta.\\

\begin{itemize}
\item {\bf H$\alpha$ line} 

The profile of H$\alpha$ is characterized by a narrow component which dominates in all our spectra of AT~2017be. Its profile is usually well reproduced by a simple Lorentzian fit. We first fitted the H$\alpha$ profile with a single Lorentzian function using standard {\sc iraf} tasks, and measured the full-width-at-half-maximum (FWHM) and flux values. In order to account for the instrumental limitations, we corrected the FWHM values for the spectral resolution ($width$ = $\sqrt{FHWM^2 - resolution^2}$), and finally we computed the H$\alpha$ FWHM velocities (v$_{FWHM}$). The resulting velocities and luminosities are reported in Table~\ref{spectra_para}, and the H$\alpha$ luminosity evolution is plotted in Figure~\ref{FWHM}.  

The velocity ranges from  $v_\mathrm{FWHM}$ $\sim$ 300 to $\sim$ 850 km~s$^{-1} $ in a rather erratic way. However, we should note that in most spectra the FWHM of H$\alpha$ is close to the instrumental resolution, hence these values are unreliable for determining the velocity of the emitting H-rich material. From our highest quality and best resolution GTC+OSIRIS spectra, we infer $v_\mathrm{FWHM}$ $\sim$ 820 km~s$^{-1} $ and $v_\mathrm{FWHM}$$\sim$ 730 km~s$^{-1}$ at phases $+30.2$ and $+92.1$ days respectively, which we regard as our most reliable values for the line velocity. During the first $\sim$ 30 days, the H$\alpha$ luminosity decreases from $\sim$ 13$\times$10$^{37}$erg~s$^{-1}$ to  $\sim$ 3$\times$10$^{37}$erg~s$^{-1}$. Later on, it settles onto slightly higher values, around 4-6$\times$10$^{37}$erg~s$^{-1}$. It is unclear if this apparent H$\alpha$ luminosity increase at later phases is real or is rather due to systematic errors (such as slit losses). However, as the H$\alpha$ luminosity increases by nearly a factor of two, we tentatively attribute it to a progressive strengthening of the interaction of ejecta with CSM.

Unfortunately, the modest resolution of our spectra is insufficient to reveal very narrow components in the H$\alpha$ profile. The only exception is the spectrum at $+30.2$ days, that can resolve features down to a limit of 160 km s$^{-1}$. This spectrum will be discussed separately later on (Section \ref{spec30}). The evolution of the H$\alpha$ profile during the three months of spectral coverage is shown in Figure~\ref{lineprofiles} (left panel). We note that there is no significant change in the wavelength position of the H$\alpha$ emission peak. 

\item {\bf Ca {\sc ii} lines} 

The evolution of the Ca~{\sc ii} NIR triplet line profiles in the wavelength window 8300-8800 \AA~for AT~2017be is shown in Figure \ref{lineprofiles} (right panel). In all spectra we mark the individual positions of Ca {\sc ii} triplet lines at 8498 \AA, 8542 \AA, and 8662 \AA. These are produced through radiative de-excitation from the upper $4p ^{2}P_{1/2,3/2}$ levels to the lower metastable $3d^{2}D_{3/2,5/2}$ levels \citep{1997A&AS..124..359M}. This triplet is a very common feature observed in the spectra of many types of transients. In our low-resolution spectra, the 8498 \AA~and 8542 \AA~members of the Ca {\sc ii} triplet are usually blended. In order to estimate the FWHM velocity of the Ca {\sc ii} lines, we considered the two spectra with the highest S/N: the NOT (phases $+0.2$ days) and GTC (phases $+92.1$ days) spectra. These spectra exhibit narrow features with velocity of 700-800 km~s$^{-1}$, which are consistent with the values obtained for H$\alpha$. This is consistent with the Ca {\sc ii} triplet arising from the same region as the main H$\alpha$ emission. On the other hand, the Ca~{\sc ii} H\&K absorptions are also clearly discernible in some epochs (the spectra at +0.2 days, +6.4 days, +12.1 days, +20.1 days and 33.2 days in Figure~\ref{all_spectra}), with a velocity of $\sim$ 180 km s$^{-1}$ (as measured from two minima of the Ca~{\sc ii} H\&K lines).

The [Ca~{\sc ii}] doublet lines at 7291,7324 \AA~are always visible in our spectra, and appear to be slightly narrower than H$\alpha$, hence in most cases these lines are unresolved. The [Ca~{\sc ii}] doublet is unblended in all our spectra, although the two components show different relative intensities and widths. The intensity ratio of the two [Ca~{\sc ii}] components ($\lambda$ 7291 vs. $\lambda$ 7328) slightly increases from $\sim$ 0.9 in our early spectra to $\sim$ 1.3 in our late spectra. However, we remark that the uncertainty in the flux ratio measurements is on the order of 20 per cent. 

As in most spectra the [Ca {\sc ii}] lines are unresolved, we consider only the two moderate resolution GTC spectra (at phases $+30.2$ days and $+92.1$ days, respectively). Also in these spectra, however, the FWHMs of the [Ca~{\sc ii}] doublet are close to the spectral resolution limit, therefore any velocity measurement is affected by a significant uncertainty. For the two spectra, we obtain $\sim$ 115 km~s$^{-1}$ and $\sim$ 250 km~s$^{-1}$ for the individual components of the doublet. The [Ca~{\sc ii}] doublet lines are produced in a relatively low density gas, from radiative de-excitation related to the metastable $3d^{2}D$ level \citep{1994ApJ...420..268C}. A major implication of the above analysis is that the [Ca~{\sc ii}] and Ca~{\sc ii} H\&K lines form in a different region as the Ca~{\sc ii} NIR triplet (and H$\alpha$). In particular, [Ca~{\sc ii}] and Ca~{\sc ii} H\&K lines are likely forming in a slowly expanding (100-200  km s$^{-1}$) circumstellar shell, while the bulk of the Ca~{\sc ii} NIR triplet and H$\alpha$ emission comes from faster-moving ejected material.

\end{itemize}

\subsection{Constraining the CSM using the +30 days spectrum} \label{spec30}

\begin{figure*}
\centering
{\subfigure[Evolution of the H$\alpha$ line profiles]{\includegraphics[width=0.45\textwidth]{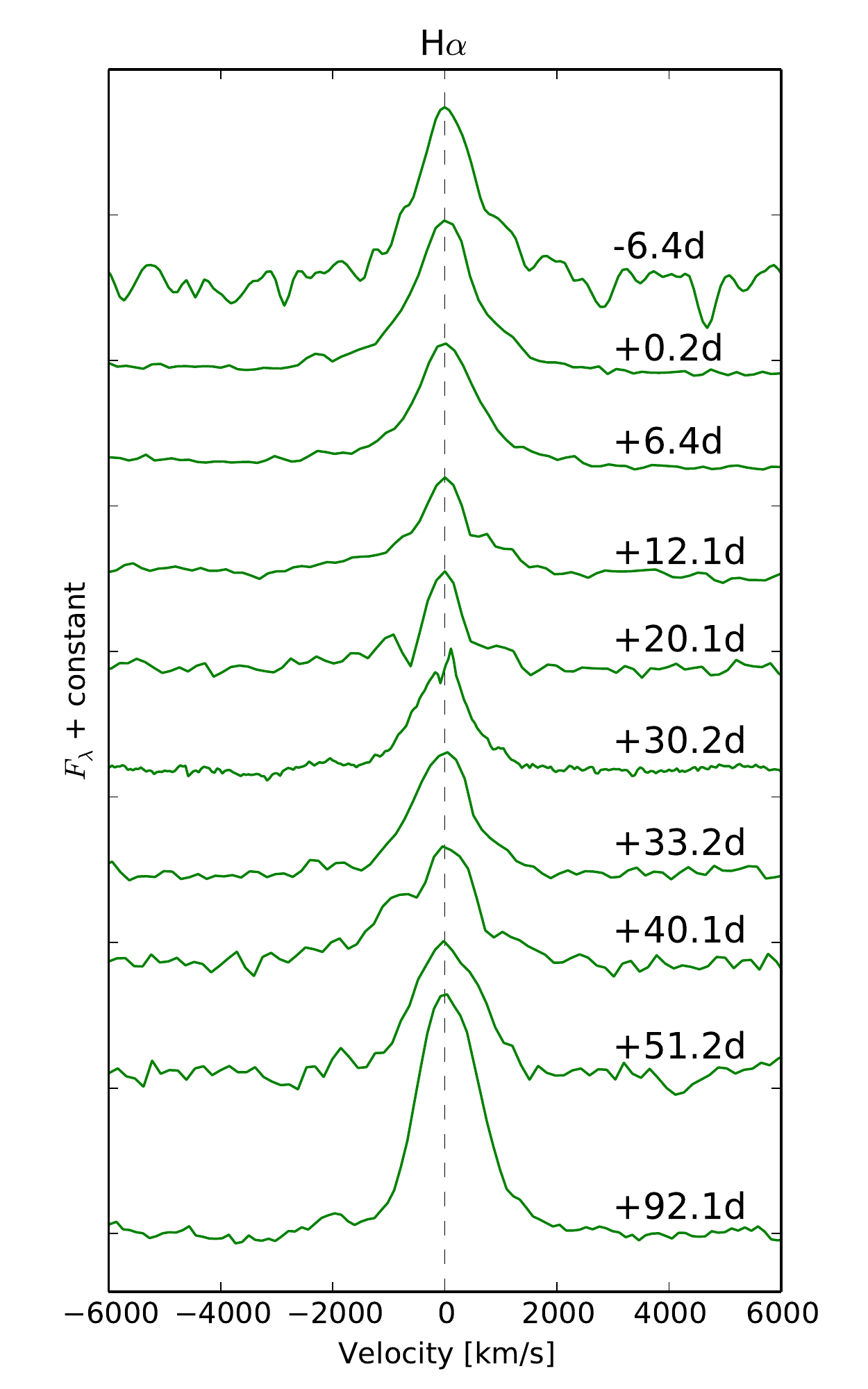}}
\mbox{\hspace{0.1cm}} 
\subfigure[Evolution of the Ca {\sc ii} line profiles]{\includegraphics[width=0.45\textwidth]{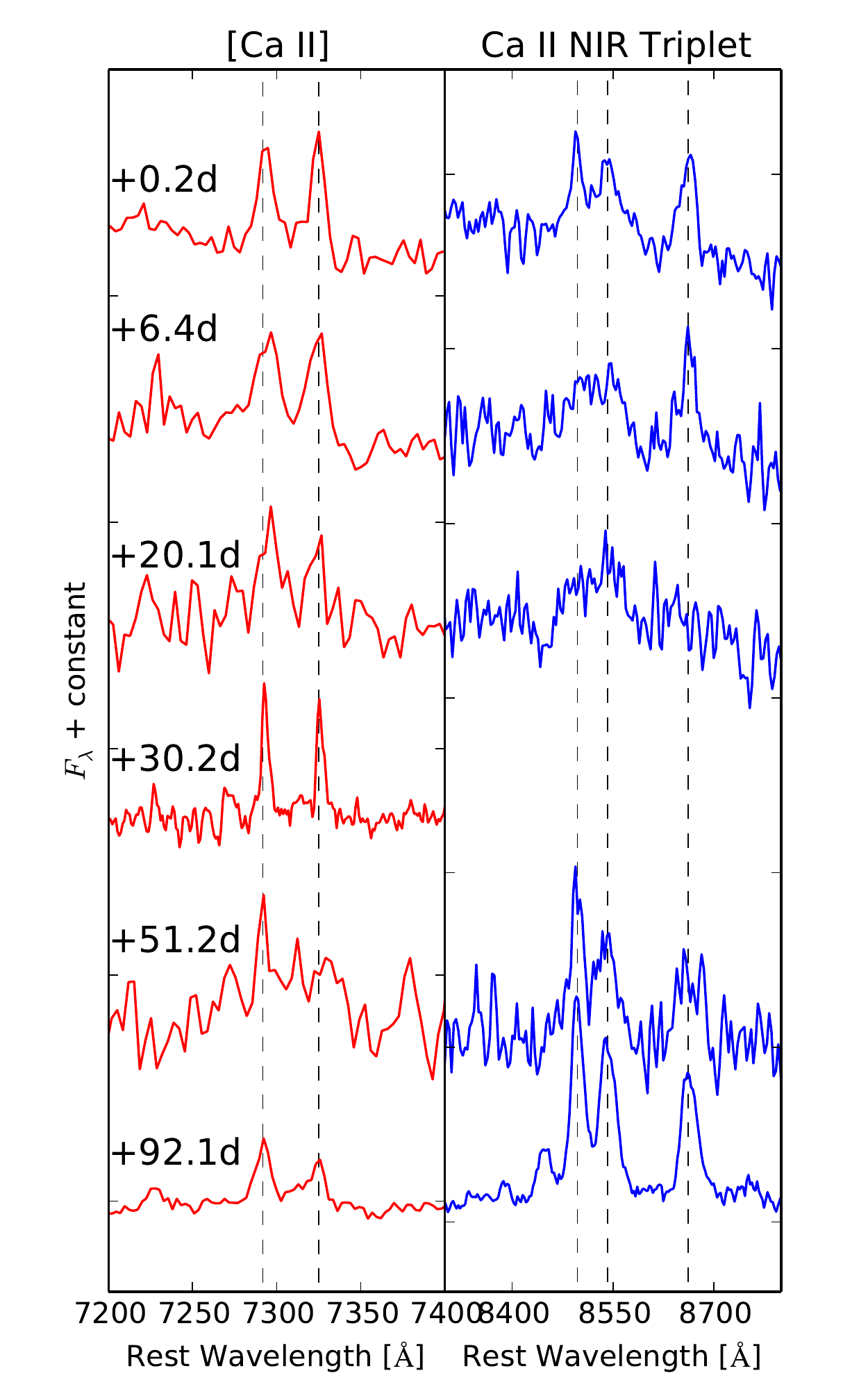}}}
\mbox{\hspace{0.1cm}}
\caption{Evolution of the H$\alpha$ (left panel), [Ca {\sc ii}] doublet and Ca {\sc ii} NIR triplet (right panel) line profiles. The vertical black dashed lines mark the rest-wavelength position of these emission lines.} 
\label{lineprofiles}
\end{figure*}

\begin{figure} 
\includegraphics[width=1.0\columnwidth]{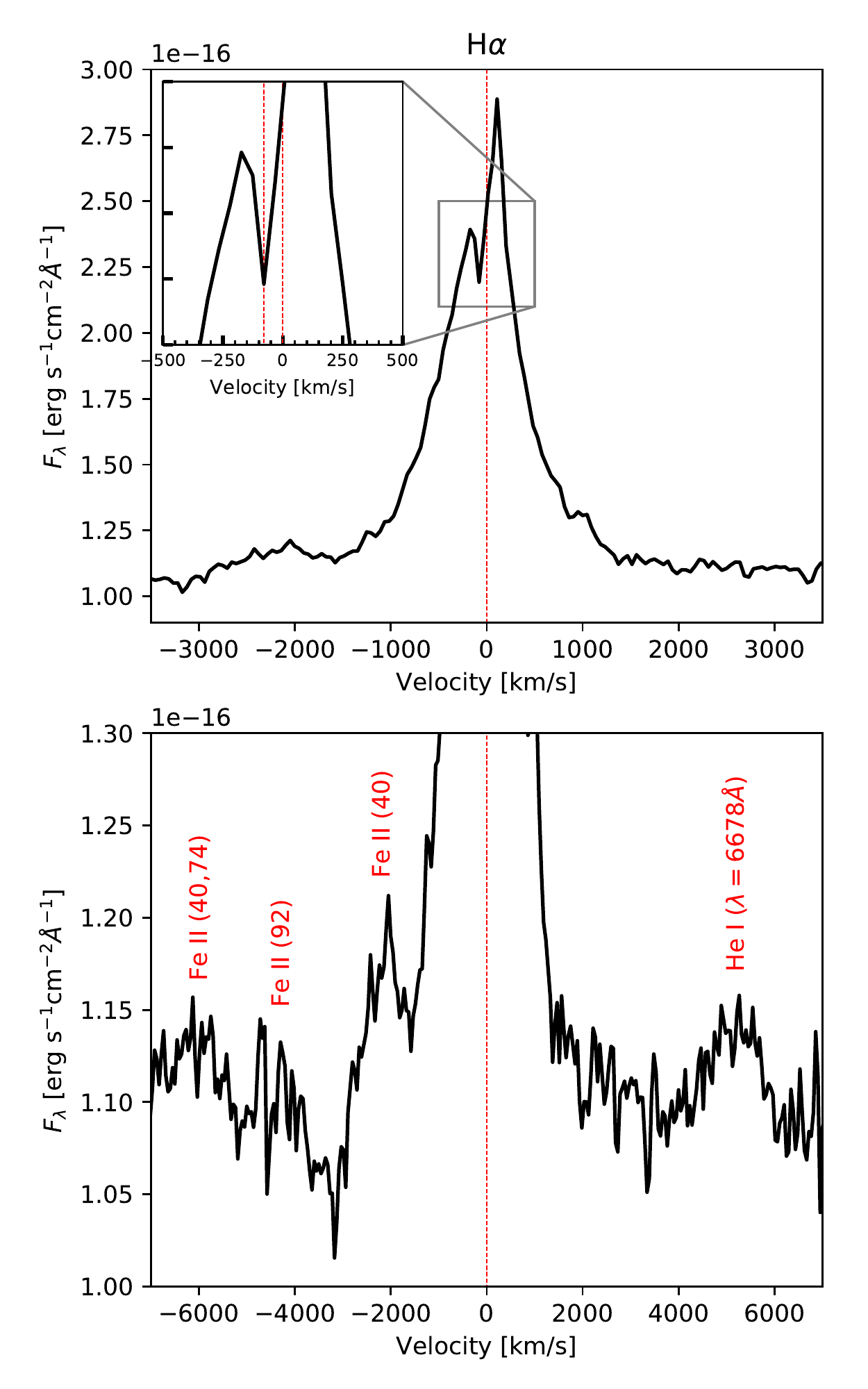}
\caption{The  profile of H$\alpha$ in our highest resolution spectrum (GTC, phase 30.2 d) in velocity space. The vertical red dashed line marks the rest-velocity position of H$\alpha$. Lower contrast Fe {\sc ii} lines and He I $\lambda$6678 are also identified in the lower panel.} 
\label{Hdetails}
\end{figure}

The  GTC+OSIRIS spectrum at phase +30.2 days (resolution  $v_\mathrm{FWHM}$ $\approx$ 160 km s$^{-1}$) reveals that the H$\alpha$ line exhibits a more complex profile than what appears from the inspection of other lower resolution spectra shown in Figure~\ref{lineprofiles}. Note, however, that some bumps likely due to blends of Fe {\sc ii} lines (see Fig. \ref{Hdetails}), are visible slightly above the flux level  of the continuum. Most of these features were observed in SN~2008S \citep[][see their figure 9]{2009MNRAS.398.1041B} and classified as Fe {\sc ii} lines (multiplets 40, 74, 92). Hence, they are not weak H$\alpha$ components. In particular the bump peaking (in the Doppler-corrected spectrum) at 6518 \AA~(v = 2100 km s$^{-1}$) is inconsistent with a blue-shifted tail of H$\alpha$, whose profile is, in fact, well-fitted by a single Lorentzian emission component with   $v_\mathrm{FWHM}$ $\sim$ 820 km~s$^{-1} $. Superimposed on the Lorentizian component, we note a very narrow P-Cygni feature (see Figure \ref{Hdetails}), which is marginally resolved, and that can be attributed to a slow-moving circumstellar shell. From the position of the narrow P-Cygni absorption, we estimate an upper limit velocity of $\sim$ 70 km~s$^{-1}$ for this H-rich shell. We note that this velocity is barely consistent with the  $v_\mathrm{FWHM}$ that we infer for the  [Ca {\sc ii}] doublet lines in the same spectrum. In fact, the average velocity obtained after instrumental resolution correction  for the two lines of the doublet is  $v_\mathrm{FWHM}$ $\sim$ 115 $\pm$ 8 km~s$^{-1}$ (the error on the line fit is only accounted for here). It likely suggests that the narrow H$\alpha$ P-Cygni and the [Ca {\sc ii}] doublet are both produced in slow-moving material ($\leq$ 70-115 km s$^{-1}$) expelled by the progenitor star years before the explosion. This modest wind velocity allows us to disfavour an LBV-like outburst as the origin for this mass-loss event, and is more consistent with CSM produced by a RSG or a super-AGB star.

\citet[]{2009ApJ...699.1850B} investigated the environmental properties of the archetypal ILRT NGC~300 OT2008-1 through high-resolution spectroscopy. 
They discussed the complex profile of prominent spectral features such as the Balmer lines, the [Ca {\sc ii}] doublet and Ca {\sc ii} NIR triplet (visible in emission) and Ca {\sc ii} H$\&$K (in absorption). These high-quality spectra of NGC 300 OT2008-1 showed also weaker features of He~{\sc i}, [O~{\sc i}] $\lambda$6300, $\lambda$6364, O~{\sc i} $\lambda$7774 and $\lambda$8446, along with Fe~{\sc ii} multiplets. The H$\alpha$ emission line was symmetric, but accompanied by narrow absorption features superimposed on the main emission, one (weaker) blue-shifted by $-$130 km s$^{-1}$, and another more prominent red-shifted by 30 km s$^{-1}$. A similar narrow P-Cygni absorption is visible in our GTC+OSIRIS spectrum (blue-shifted by $-$70 km s$^{-1}$). As in the case of NGC 300 OT2008-1, the [Ca {\sc ii}] doublet has a strongly asymmetric profile without a blue wing, while the Ca {\sc ii} NIR triplet has a broad absorption feature, supporting the existence of multiple line components. The features of \citeauthor{2009ApJ...699.1850B}'s high resolution spectra suggest that NGC 300 OT2008-1 has a complex circumstellar environment, marked by outflows and  inflows from different gas components, including the wind from a possible companion star (a Wolf-Rayet star or a blue supergiant).

\begin{table*}
\caption{Main parameters inferred from the spectra of AT~2017be.}
\label{spectra_para}
\begin{tabular}{@{}ccccccc@{}}
\hline
Date & MJD & Phase$^a$  &  FWHM   H$\alpha$   & Velocity H$\alpha$  &        Luminosity H$\alpha$$^d$      &           Temperature    \\ 
        &          & (d)                &          ($\ang$)           &    (km~s$^{-1} $)     &   (10$^{37}$erg~s$^{-1}$)   &     (K)       \\ 
\hline
20170110$^c$ & 57763.4 & $-6.4$    &  19.9              & 570 &  13.0     &   7650 $\pm$ 1250  \\
20170117$^c$ & 57770.0 & $+0.2$   &  19.8              &  590  &  10.2     &   6650 $\pm$ 400     \\
20170123$^c$ & 57776.2 & $+6.4$   &  20.8              & 480  &  7.3       &   6550 $\pm$ 300   \\
20170128$^c$ & 57781.9 & $+12.1$ &  18.6              & 500   &  4.1       &   6550 $\pm$ 550    \\
20170205$^c$ & 57790.0 & $+20.1$ &   19.2$^b$    & 310   &  3.7       &   5150 $\pm$ 500    \\
20170216         & 57800.0 & $+30.2$ &   18.4             &  820         &  3.0      &   5000 $\pm$ 900      \\
20170219$^c$ & 57803.0 & $+33.2$ &   19.3             & 610   &  3.8      &   5250 $\pm$ 350    \\
20170225$^c$ & 57809.9 & $+40.1$ &    24.9            & 740   &  5.3      &   5400 $\pm$ 500     \\
20170308$^c$ & 57821.0 & $+51.2$ &    22.0            & 580   &  4.3      &   5650 $\pm$ 750     \\
20170418          & 57861.9 & $+92.1$ &    17.9            & 730           &  5.9      &   5100 $\pm$ 350    \\

\hline
\end{tabular}

\medskip

$a$ Phases are relative to $r$-band maximum light, on MJD = 57769.8 $\pm$ 0.1.\\ 
$b$ There is a hot pixel close to the H$\alpha$ position, affecting the recovered line profile and making the above measurement uncertain.\\
$c$ These measurements are close to the instrumental resolution, hence after correcting for instrumental resolution the resulting velocity has a large uncertainty.\\
$d$ We assume conservative errors on the luminosity of about 20 per cent.

\end{table*}

\section{Progenitor analysis}\label{progenitor}

In the attempt to detect (or place stringent detection limits) to the quiescent progenitor star, we searched pre-explosion images of the field of AT~2017be in ground based and space telescope archives. Unfortunately while the host galaxy is relatively nearby, very few deep images are available in the archives. The most constraining image is a {\sl HST + NICMOS} frame taken on 2002 October 14, using the NIC3 camera and the F160W filter ($\sim$ $H$ band). A narrow band NIR image taken at the same epoch does not place useful magnitude limits to the progenitor. 

\begin{figure*} 
\centering
\includegraphics[width=2\columnwidth]{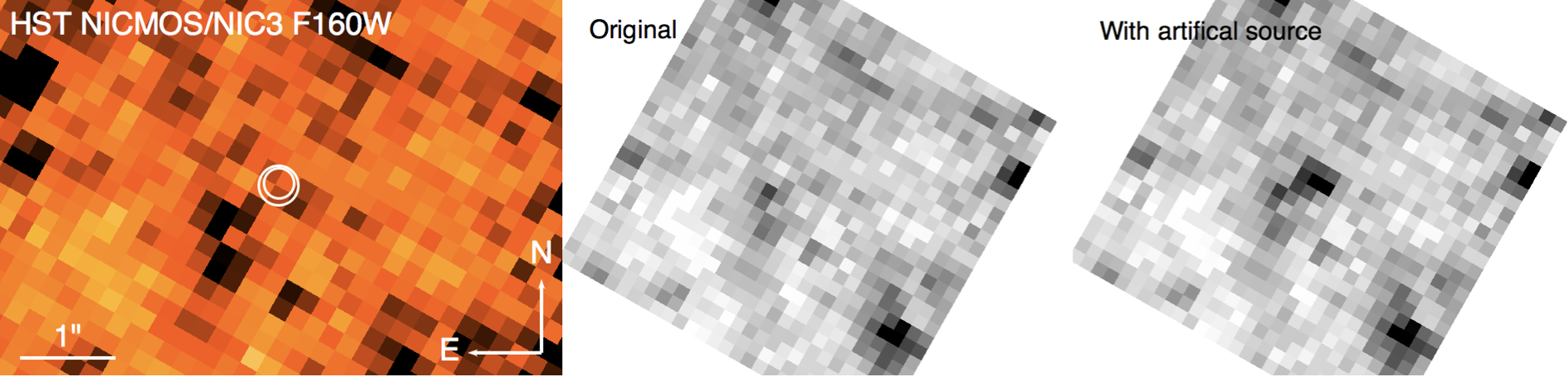}
\caption{A section of the archival HST+NICMOS F160W image of the site of AT~2017be. The circles mark the location of AT~2017be as determined from $i$- and $H$-band images of the SN; the radii of the circles correspond to 3x the RMS scatter in the astrometric transformation (left).The HST+NICMOS F160W image of the site of AT~2017be (middle), and the same region with an artificial source at 3$\sigma$ significance (right)}
\label{HST}
\end{figure*}

To identify the position of AT~2017be in the NICMOS data, we aligned the image to good seeing images taken at the NOT. We aligned the NICMOS data to both an $i$-band image taken on 2017 April 21, and a $H$-band image taken on April 8, using 10 sources in the former and 7 in the latter that were also present in the NICMOS frame. Rotation, translation and a scale factor between each pair of images were fitted for, and the resulting transformation had an RMS scatter of 69~mas and 53~mas when using either $i$ band or $H$ band, respectively, as our reference image. In both cases, we obtained an almost identical position for AT~2017be on the NICMOS image (see the left panel of Figure~\ref{HST}). 

To determine a limiting magnitude for the progenitor of AT~2017be, we performed artificial star tests on the NICMOS data (see the middle and right panels of Figure~\ref{HST}). A model point spread function was created from sources in the field, and then scaled in flux and inserted at the SN position. The 3$\sigma$ limiting magnitude was taken to be the value at which the source was recovered with a photometric uncertainty of 0.2 mag. We determine a flux limit for the progenitor with this technique of $F<2.04\times\, 10^{-18}$ erg~s$^{-1}$~cm$^{-2}$~\AA$^{-1}$, or a limiting magnitude in the ABmag system of $F160W>20.8$~mag. 

\begin{figure*} 
\centering
\includegraphics[width=2\columnwidth]{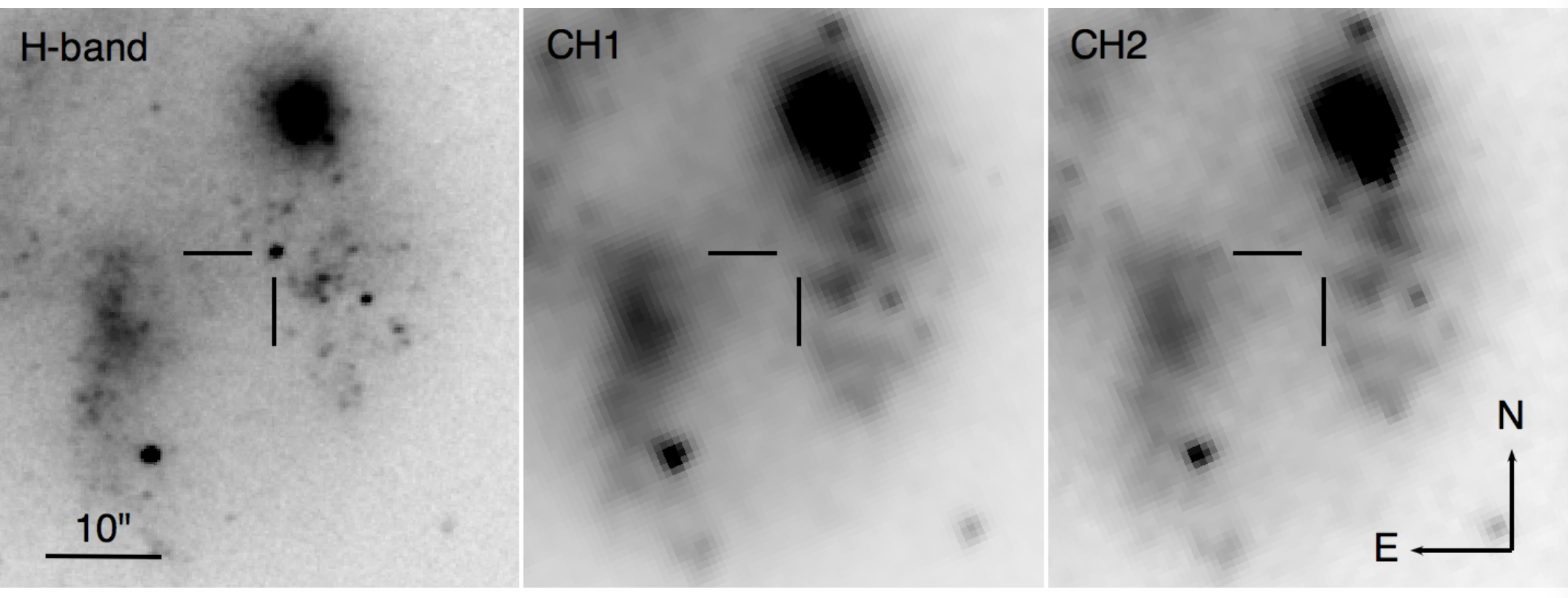}
\caption{$H$ band post-explosion and Spitzer IRAC CH1 (3.6$\upmu$m) and CH2 (4.5$\upmu$m) pre-explosion images of the site of AT~2017be. The position of the transient is indicated with tick marks in all panels.}
\label{mir}
\end{figure*}

We used the same techniques to determine limiting magnitudes for the Spitzer IRAC images (2004-04-01.5 UT, see the Figure~\ref{mir}) as were used for the NICMOS data. Limits of M$_{3.6\upmu}$ > 15.80 mag and M$_{4.5\upmu}$ > 15.82 mag (Vega mag) were found in IRAC CH1 and CH2 respectively. 

The limits we can place on the progenitor of AT~2017be are unfortunately not particularly constraining. The implied absolute magnitude of the progenitor given a distance modulus of $\upmu$ = 29.47 mag is fainter than $H$ = $-$10.1 mag (converting to Vegamag). For comparison, the progenitor of SN 2008S, which was detected in the MIR with Spitzer, had a $K$-band limit of $> -$10.8 mag \citep{2009MNRAS.398.1041B}. 

\citet{2017arXiv171110501A} performed an independent analysis on pre-explosion archive images, although they did not describe in depth the methods used to infer their limits. While the $H$-band limit inferred by Adams et al. from the HST + NICMOS data, is similar to ours, their IRAC limits are significantly deeper \cite[][almost 1.5 mag discrepancy for CH1 and CH2, respectively; see their Table 4]{2017arXiv171110501A}. We note that the putative progenitor was blended with an IR-luminous extended source, and this can possibly explain the discrepancy. However, as it is unclear from \citeauthor{2017arXiv171110501A} exactly what data were used in their analysis, and how their limiting magnitudes were determined, we cannot investigate this discrepancy further.


\section{Discussion and Summary} \label{discussion}
\subsection{Comparison with other ILRTs} 

In recent years, the number of discovered ILRTs has grown significantly, and includes such events as SN~2008S \citep{2009MNRAS.398.1041B}, NGC~300~OT2008-1 \citep{2011ApJ...743..118H}, M85~OT2006-1 \citep[][]{2007Natur.447..458K} and PTF10fqs \citep{2011ApJ...730..134K}. The latest addition is AT~2017be, whose absolute magnitude at peak lies in the same range ($-10 > M_{V} > -15$ mag) as several putative classes of ILOTs.
More specifically, AT~2017be exhibits the characteristic properties an ILRT, having a relatively slow spectro-photometric evolution, along with a lightcurve that resembles those of Type II-P or Type II-L SNe. In addition, the spectra of AT~2017be  are dominated by prominent and narrow H emission lines ($v_\mathrm{FWHM}$ < 1000 km~s$^{-1}$), weak Fe~{\sc ii} features, Ca~{\sc ii} H$\&$K, and the Ca~{\sc ii} NIR triplet, similar to other ILRTs. Most importantly, the [Ca~{\sc ii}] doublet (see Figure \ref{comparison}, right), which is a typical marker of ILRTs, is visible in all our spectra.

The lightcurve of AT~2017be is very similar to that of PTF10fqs, and shares some similarity with M85~OT2006-1. Their lightcurves show a sort of short-duration plateau. The colour evolution of AT~2017be in the optical bands matches that of NGC 300 OT2008-1, especially the $B-V$ and $R-I/r-i$ colours. The $J-K$ colour of AT~2017be, SN~2008S, NGC 300~OT2008-1 and M85~OT2006-1 becomes redder with time. The Sloan $r$-band  absolute magnitude at maximum of AT~2017be is $M_{r}$ = $-$11.98 $\pm 0.09$ $\rm{mag}$, and is almost equal to that of PTF10fqs. The constraint on the $^{56}$Ni mass inferred from the late phase quasi-bolometric lightcurve, implies that very little $^{56}$Ni was ejected in the explosion ($\sim$ 8 $\times$ 10$^{-4}$\msun~or less), which is less than that estimated by \citeauthor{2009MNRAS.398.1041B} for SN~2008S. 

A common characteristic of ILRTs is their MIR luminous pre-explosion progenitors, with the quiescent star being embedded in a dusty CSM \cite[][see their figure 17]{2009MNRAS.398.1041B}. The best-studied cases indicate moderate-mass (8-12 \msun) progenitors, although in some cases the studies point to slightly lower-mass stars \citep[e.g. M85 OT2006-1,][]{2007Natur.449E...1P, 2007ApJ...659.1536R, 2008ApJ...674..447O}. Unfortunately, we cannot set a stringent constraints on the AT~2017be progenitor star in pre-explosion archive images.\\

\subsection{Plausible scenarios for ILRTs and conclusions} \label{conclusions}
The nature of ILRTs is  debated. Several scenarios have been  proposed in the attempt to interpret their peculiar observables. For SN~2008S, a scenario invoking a weak ECSN explosion of a super-AGB star embedded in an optical thick dust envelope has been proposed \citep{2008ApJ...681L...9P, 2009MNRAS.398.1041B, 2009ApJ...705.1364T}. Alternatively, the LBV-like super-Eddington outburst of a highly obscured, massive progenitor ($\sim$ 20\msun) has been suggested by \citet{2009ApJ...697L..49S}. Another well-studied object is NGC 300 OT2008-1, and which is still consistent with both an ECSN  \citep{{2008ApJ...681L...9P},2009ApJ...705.1364T} or an LBV-like eruption \citep{2011MNRAS.415..773S}. 

\citet{2009ApJ...699.1850B} investigated NGC~300~OT2008-1 with  high-resolution spectroscopy, and proposed that the observed outburst was produced by a $\sim$ 10-20\msun, relatively compact progenitor (a blue supergiant or a Wolf-Rayet star), obscured by a dusty cocoon. An alternative explanation involved interaction in a binary system. \citet{2009ApJ...695L.154B} proposed NGC 300 OT2008-1~to be an eruption of a  dust-enshrouded, IR-luminous and relatively massive star ($\sim$ 10-15\msun), likely an OH/IR source, which  began its evolution on a blue loop toward higher temperatures. Both SN~2008S and NGC 300 OT2008-1 are slightly more luminous than other ILRTs, and have lightcurves similar to those of Type II-L SNe. From an observational point of view, ILRTs appear to be scaled-down analogs of Type II-P or Type II-L SNe.

Therefore, we now discuss three possible scenarios (stellar mergers, LBV-like eruption, and EC SN explosions) for ILRTs.

\begin{enumerate}

\item The most extreme outcome of binary interaction is a stellar merger. The discovery of V1309 Sco \citep{2017ASPC..510..401B} provided strong support to the idea that Luminous Red Novae (LRNe) likely result from a merging event \citep[see, e.g.,][]{2010A&A...516A.108M,2011A&A...528A.114T}. The typical observational features of LRNe such as V8383 Mon, V1309 Sco and V4332 Sgr are multi-peaked lightcurves, low ejecta velocities ($\sim$ 100-300 km~s$^{-1}$), no  evidence of [Ca {\sc ii}] features, and very late-time spectra dominated by molecular bands (TiO, VO) in the optical domain \citep[][]{1999AJ....118.1034M,  2003MNRAS.341..785C,  2003A&A...412..767R, 2004ApJ...604L..57B, 2013CEAB...37..325B, 2014AJ....147...11M, 2015A&A...580A..34K,  2015AJ....149...17L, 2015A&A...578A..75T}. All of this makes AT~2017be quite different from LRNe, hence we agree with \citeauthor{2017arXiv171110501A} in ruling out a stellar merger scenario for this transient.

\item LBVs are erupting massive stars that may experience severe mass-loss. Multiple outbursts can be observed during a single eruptive phase that may lasts even decades (like the Great Eruption of $\eta$~Carinae). During these outburst, no $^{56}$Ni is expected to be ejected. In addition, the ejected gas expands at velocities of a few hundreds km~s$^{-1}$, occasionally exceeding 10$^4$ km~s$^{-1}$ during a giant eruption, which are inconsistent with our spectroscopic observations (e.g., $\sim$ 820 km~s$^{-1}$ and 730 km~s$^{-1}$ at epoch of +30.2 days and +92.1 days, respectively). An LBV outburst scenario for AT~2017be has been recently proposed by \citet{2017arXiv171110501A}. However, we do not find robust arguments to support  this, as we did not detect any luminous progenitor in pre-explosion data (in contrast with expectations from a massive LBV). A caveat here is that lower luminosity LBVs may have initial masses down to  about 20\msun, as claimed by \cite{2010AJ....139.1451S, 2016MNRAS.455.3546S, 2018arXiv180503298S}. In particular, UGC 2773-OT is a moderate-mass LBV of $\sim$ 20\msun, sharing some observational properties with ILRTs (in particular, the overall spectral evolution and the detection of the [Ca II] doublet). However, the light curve was 
totally different, showing a rise phase lasting over a decade, which is a clear signature of a long-lasting eruptive phase. AT~2017be and ILRTs have short-duration light curves, resembling that of some CC SNe, and the best-followed objects (SN~2008S and NGC~300 OT2008-1) show a late-time decline rate consistent with the $^{56}$Co decay into $^{56}$Fe, suggesting a SN explosion (see below). \\

\item \citet{2009ApJ...705L.138P} described the expected properties of EC SNe from super-AGB progenitors combining theoretical models with observational evidence. This type of CC SNe are  triggered by electron-capture in the core of a progenitor which has evolved to  reach the super-AGB phase. The threshold to trigger an EC SN explosion is determined by the competition between core growth and mass loss \citep[see, e.g.,][]{2002RvMP...74.1015W,2005ARA&A..43..435H}. If the core grows to reach the Chandrasekhar limit  \citep[i.e., $M_{CH}$$\sim$ 1.37\msun; ][]{1984ApJ...277..791N}, electron-capture reactions are triggered, allowing the super-AGB star to explode as faint ($^{56}$Ni-poor) ECSN. In contrast, if envelope mass loss is large enough, electron-capture reactions  fail, and the super-AGB star evolves to become an ONe white dwarf (WD). Although the ECSN mechanism in super-AGB is a plausible explanation, several uncertainties still exist. Nonetheless, as a major support to the ECSN scenario for ILRTs, \citet{2016MNRAS.460.1645A} obtained late-time Hubble and Spitzer Space Telescope imaging of SN 2008S and NGC~300~OT2008-1, and detected both transients at a mid-IR luminosity fainter than those of their quiescent progenitors, without a detection in the optical and NIR bands. 

\end {enumerate}

Unfortunately, in contrast with SN~2008S and NGC~300 OT2008-1, a tight constraint on the progenitor of AT~2017be does not exist. However, we favour an ECSN scenario on the basis of the following supporting observational arguments: 
\begin{itemize}
\item The lightcurves of ILRTs resemble sub-luminous counterparts of Type II-P or II-L SNe.
\item The expansion velocity of the ejecta inferred from broader resolved H and Ca {\sc ii} lines of AT~2017be is significantly lower than those of canonical CC SNe, but only slightly slower than those observed in some faint Type II SNe \citep{2004MNRAS.347...74P, 2014MNRAS.439.2873S}. The wind velocity expected in the super-AGB stars is low, $\sim$ 100 km~s$^{-1}$ \citep{1979BAAS...11Q.724W}, or even less. This agrees with the velocities inferred for the slow-moving CSM in the highest resolution spectra of NGC~300~OT2008-1 \citep{2009ApJ...699.1850B}, and the GTC spectrum of AT~2017be at +30.2 d. 
\item AT~2017be was photometrically monitored until very late epochs. Although already below the detection threshold in the other monitoring bands, in the $K$-band it was visible up to $\sim$ 350 days from maximum. The late time photometric data allow us to constrain to about 8 $\times$ 10$^{-4}$\msun~for the ejected $^{56}$Ni mass. This $^{56}$Ni limit is over a factor of one lower than the amount estimated for other ILRTs ($\lesssim$ $10^{-3}$\msun). The above limit is still consistent with the predictions for an EC SN explosion \citep{2009MNRAS.398.1041B,2009ApJ...705L.138P}. However, on the other hand, we can not exclude a totally $^{56}$Ni-free outburst.  
\item For a few ILRTs, the progenitor stars were recovered in pre-explosion archive images, and in all cases they were moderate mass stars  (in the mass range $\sim$ 8-12\msun) embedded in a dusty environment, consistent with super-AGB star scenario.
\item \citet{2016MNRAS.460.1645A} reported the mid-IR fading of some ILRTs at very late phases to be more than 15 times fainter than the luminosity of the quiescent progenitor detection. This is a crucial argument to support a terminal SN explosion of ILRT progenitor stars. A deep IR limit with James Webb Space Telescope (JWST) can determine if there is a surviving star following AT~2017be.  
\end{itemize}

It is evident that high-resolution spectroscopy at early phases, the availability of deep pre-discovery images in the optical and IR archives (ideally with high spatial resolution), and deep late-time observations in the IR domain are crucial for better constraining the explosion scenario of AT~2017be and other ILRTs. Future surveys in the optical bands (e.g. with the Large Synoptic Survey Telescope; LSST) will discover a larger number of new ILRTs, enabling us to unequivocally unveil the\citep{Quimby2011Natur.474..487Q} nature of their stellar progenitors and their explosion mechanism.\\

\newpage
\section*{Acknowledgements}
We thank the reviewer for his/her insightful comments that have improved the paper. We thank M. Ergon, Z. Kostrzewa-Rutkowska, M.~Nielsen, and G. Fedorets for their observations obtained on 2017 February 18 and June 18, 2017 March 9, 2017 February 5,  and 2017 September 27 respectively. We are particularly grateful to S. Valenti for his support with LCO. Y-Z Cai thanks the useful discussion with M. Turatto, L. Tartaglia, A. Reguitti, S. Yang, and A. Fiore. 

Based on observations made with: \\
-The Nordic Optical Telescope (NOT), operated by the NOT Scientific Association at the Spanish Observatorio del Roque de los Muchachos of the Instituto de Astrofisica de Canarias. \\
-The LCO 1.0-m telescope at McDonald Observatory (Texas, USA) equipped with an 1-m Sinistro (fl05) camera and 2.0-m telescope at Haleakala Observatory (Hawaii, USA) equipped with the FLOYDS spectrograph. \\
-The Gran Telescopio Canarias (GTC) operated on the island of La Palma at the Spanish Observatorio del Roque de los Muchachos of the Instituto de Astrofisica de Canarias. \\
-The Cima Ekar 1.82~m Telescopio Copernico of the INAF (Istituto Nazionale di Astrofisica) -- Astronomical Observatory of Padova, Italy. \\
-The Intermediate Palomar Transient Factory (iPTF) 1.2~m Samuel Oschin telescope. 

Y-Z Cai is supported by the China Scholarship Council (No. 201606040170). \\
M.F. acknowledges the support of a Royal Society - Science Foundation Ireland University Research Fellowship. \\
C.G. acknowledges support from the Carlsberg Foundation. \\ 
Support for IA was provided by NASA through the Einstein Fellowship Program, grant PF6-170148. \\
G.H., D.A.H. and C.M. are supported by the National Science Foundation (NSF) under grant AST-1313484.\\
J.H. acknowledges financial support from the Vilho, Yrj\"o and Kalle V\"ais\"al\"a Foundation of the Finnish Academy of Science and Letters.\\
This work has been supported in part by a research grant 13261 (PI M. D. Stritzinger) from the VILLUM FONDEN.

Observations from the NOT were obtained through the NUTS collaboration which is supported in part by the Instrument Centre for Danish Astrophysics (IDA). This work makes use of data from Las Cumbres Observatory and these data were taken as part of the LCO Supernova Key Project. 
Funding for SDSS-III has been provided by the Alfred P. Sloan Foundation, the Participating Institutions, the National Science Foundation, and the U.S. Department of Energy Office of Science.  The SDSS-III web site is http://www.sdss3.org/.  
This publication makes use of data products from the Two Micron All Sky Survey, which is a joint project of the University of Massachusetts and the Infrared Processing and Analysis Center/California Institute of Technology, funded by the National Aeronautics and Space Administration and the National Science Foundation.
\textsc{iraf} is distributed by the National Optical Astronomy Observatories, which are operated by the Association of Universities for Research in Astronomy, Inc., under cooperative agreement with the National Science Foundation.
This research has made use of the NASA/IPAC Extragalactic Database (NED), which is operated by the Jet Propulsion Laboratory, California Institute of Technology, under contract with the National Aeronautics and Space Administration.
We acknowledge the usage of the HyperLeda database (http://leda.univ-lyon1.fr).


\bibliographystyle{mnras}
\bibliography{impostors.bib}


\appendix

\section[]{Lightcurves of AT~2017be}
\begin{table*}
\begin{minipage}{175mm}                                  
\caption{Optical ($BVgriz$) photometric data for AT~2017be. Early unfiltered observations are also reported.}
\label{optical_bands}  
\begin{tabular}{@{}ccccccccccl@{}}
\hline
Date         &     MJD       &phase$^{a}$&  $B$(err)         &  $V$(err)         &  $g$(err)          & $r$(err)            & $i$(err)            & $z$(err)         & Instrument \\
\hline
20161229 & 57751.48 &$-$18.34&   $\cdots$            &    $\cdots$               & $\cdots$            &  >18.7$^b$    &  $\cdots$          &  $\cdots$        & LOSS$^e$     \\
20170106 & 57759.51 &$-$10.31&    $\cdots$              &    $\cdots$               & $\cdots$        &  18.71$^b$    & $\cdots$      &  $\cdots$           & LOSS$^e$     \\
20170107 & 57760.12 &$-$9.70&      $\cdots$             &    $\cdots$              & $\cdots$                     &  18.30             & $\cdots$       &  $\cdots$          & iPTF$^f$  \\
20170107 & 57760.16 &$-$9.66&  >17.8    &     $\cdots$     & $\cdots$     & $\cdots$                     &  $\cdots$                  & $\cdots$      & fl05    \\
20170107 & 57760.43 &$-$9.39&      $\cdots$              &    $\cdots$              &  $\cdots$                     &  18.347         &  $\cdots$      &  $\cdots$      & iPTF$^g$  \\
20170108 & 57761.36 &$-$8.47& 18.982(0.234) & 18.442(0.122) & 18.645(0.216) & 18.135(0.144) & 18.010(0.065) &  $\cdots$           & fl05    \\
20170108 & 57761.42 &$-$8.40& 18.970(0.097) & 18.454(0.061) & 18.660(0.118) & 18.140(0.060) & 18.011(0.047) &  $\cdots$           & fl05    \\
20170110 & 57763.34 &$-$6.48& 18.286(0.278) & 17.986(0.084) & 18.171(0.200) & 17.818(0.091) & 17.742(0.083) &  $\cdots$         & fl05   \\
20170110 & 57763.42 &$-$6.40&  >17.8         &  >17.141         & 18.193(0.217) & 17.821(0.199) & 17.744(0.115) &  $\cdots$             & fl05   \\
20170116 & 57769.20 &$-$0.63& 18.449(0.088) & 18.142(0.073) & 18.276(0.123) & 17.756(0.065) & 17.748(0.088) &  $\cdots$          & fl05    \\
20170116 & 57769.98 &+0.16& 18.496(0.031) & 18.167(0.034) & 18.326(0.027) & 17.743(0.026) & 17.792(0.031) &17.783(0.068) & ALFOSC  \\
20170120 & 57773.27 &+3.45& 18.518(0.075) & 18.046(0.055) & 18.209(0.059) & 17.835(0.050) & 17.777(0.048) & $\cdots$          & fl05   \\
20170120 & 57773.31 &+3.49& 18.538(0.083) & 18.067(0.046) & 18.319(0.048) & 17.908(0.040) & 17.766(0.053) & $\cdots$            & fl05   \\
20170123 & 57776.15 &+6.33& 18.733(0.023) & 18.090(0.047) &  $\cdots$                 &  $\cdots$                  & 17.905(0.139) & $\cdots$                & ALFOSC \\
20170123$^c$ & 57776.95 &+7.13& 18.870(0.153) & 18.112(0.090) & 18.409(0.096) & 17.914(0.083) & 17.853(0.088) &17.967(0.075) & AFOSC  \\
20170124 & 57777.28 &+7.46& 18.699(0.077) & 18.160(0.057) & 18.427(0.051) & 18.028(0.042) & 17.897(0.045) & $\cdots$                & fl05  \\
20170127 & 57780.33 &+10.51& 18.852(0.087) & 18.229(0.057) & 18.634(0.061) & 18.121(0.045) & 18.037(0.052) &17.938(0.129) & fl05    \\
20170128 & 57781.87 &+12.05& 18.989(0.074) & 18.330(0.043) & 18.665(0.045) & 18.144(0.041) & 18.028(0.051) &17.954(0.057) & ALFOSC  \\
20170131 & 57784.27 &+14.45& 19.088(0.102) & 18.358(0.065) & 18.837(0.066) & 18.207(0.056) & 18.078(0.052) &17.927(0.106) & fl05 \\
20170204 & 57788.25 &+18.43& 19.194(0.123) & 18.427(0.066) & 18.750(0.071) & 18.208(0.058) & 18.087(0.068) &17.996(0.122) & fl05   \\
20170205$^{c,d}$ & 57789.95 &+20.13& 19.297(0.043) & 18.521(0.031) & 18.809(0.043) & 18.229(0.038) & 18.058(0.031) &17.966(0.075) & ALFOSC \\
20170209 & 57793.37 &+23.55& 19.427(0.207) & 18.435(0.089) & 18.881(0.069) & 18.243(0.067) & 18.086(0.071) &17.974(0.150) & fl05  \\
20170217 & 57801.22 &+31.40& 19.540(0.280) & 18.668(0.106) & 19.084(0.166) & 18.368(0.058) & 18.156(0.041) &17.988(0.084) & fl05  \\
20170219 & 57803.07 &+33.25& 19.569(0.046) & 18.627(0.099) & 19.175(0.032) & 18.451(0.021) & 18.185(0.036) &18.058(0.040) & ALFOSC  \\
20170221 & 57805.26 &+35.44& 19.598(0.369) & 18.774(0.245) & 19.214(0.399) & 18.493(0.062) & 18.194(0.038) &18.052(0.064) & fl05  \\
20170225 & 57809.14 &+39.32& 19.772(0.202) & 18.855(0.196) & 19.275(0.192) & 18.511(0.121) & 18.263(0.064) &18.073(0.108) & fl05  \\
20170225 & 57809.91 &+40.09& 19.750(0.047) & 18.833(0.032) & 19.288(0.031) & 18.513(0.027) & 18.269(0.016) &18.109(0.042) & ALFOSC   \\
20170302 & 57814.28 &+44.46&   $\cdots$       & 18.958(0.193) & 19.364(0.107) & 18.721(0.189) & 18.389(0.133) & $\cdots$          & fl05  \\
20170307 & 57819.20 &+49.38& 19.962(0.132) & 18.946(0.060) & 19.263(0.068) & 18.570(0.045) & 18.372(0.043) &18.112(0.085) & fl05  \\
20170309 & 57821.02 &+51.20& 19.950(0.076) & 18.923(0.047) & 19.317(0.045) & 18.653(0.028) & 18.365(0.020) &18.153(0.040) & ALFOSC \\
20170311 & 57823.26 &+53.44& 20.019(0.222) & 18.977(0.172) & 19.359(0.132) & 18.621(0.136) & 18.431(0.134) &17.992(0.307) & fl05  \\
20170315 & 57827.22 &+57.40& 20.024(0.190) & 18.991(0.079) & 19.387(0.058) & 18.694(0.065) & 18.450(0.053) &18.132(0.112) & fl05     \\
20170319 & 57831.19 &+61.37& 20.052(0.180) & 19.028(0.085) & 19.406(0.084) & 18.710(0.056) & 18.488(0.071) &18.157(0.083) & fl05     \\
20170327 & 57839.14 &+69.32& 20.085(0.092) & 19.129(0.055) & 19.595(0.055) & 18.809(0.085) & 18.551(0.104) &18.360(0.136) & fl05    \\
20170328 & 57840.87 &+71.05& 20.119(0.093) & 19.172(0.098) & 19.604(0.176) & 18.879(0.059) & 18.623(0.046) &18.368(0.082) & AFOSC    \\
20170403 & 57846.10 &+76.28& 20.122(0.191) & 19.329(0.119) & 19.816(0.077) & 19.019(0.057) & 18.785(0.051) &18.336(0.121) & fl05    \\
20170407 & 57850.94 &+81.12& 20.216(0.079) & 19.416(0.042) & 19.993(0.084) & 19.147(0.052) & 18.831(0.040) &18.500(0.038) & ALFOSC   \\
20170409 & 57852.22 &+82.40& 20.302(0.161) & 19.414(0.127) & $\cdots$                     & 19.162(0.085) & 18.817(0.072) &18.510(0.112) & fl05    \\
20170411 & 57854.21 &+84.39& >19.9         & 19.438(0.295) & 19.918(0.312)  &   >19.2        & $\cdots$           & $\cdots$                 & fl05    \\
20170414 & 57857.14 &+87.32& 20.349(0.112) & 19.634(0.093) &  $\cdots$     &   $\cdots$    &   $\cdots$        & $\cdots$               & fl05    \\
20170417 & 57860.11 &+90.29& 20.432(0.548) & 19.726(0.150) & 20.259(0.065) & 19.262(0.046) & 18.979(0.043) &18.590(0.095) & fl05    \\
20170421 & 57864.89 &+95.07& 20.860(0.118) & 19.835(0.113) & 20.287(0.051) & 19.411(0.059) & 19.131(0.078) &18.756(0.064) & ALFOSC   \\
20170502 & 57875.87 &+106.05& 21.140(0.090) & 20.235(0.056) & 20.567(0.067) & 19.599(0.045) & 19.318(0.045) &18.886(0.054) & ALFOSC   \\
20170505 & 57878.12 &+108.30& >20.9          & 20.387(0.470) & 20.704(0.202) & 19.719(0.109) & 19.344(0.105) &18.861(0.149) & fl05    \\
20170515 & 57888.13 &+118.31& 21.303(0.469) & 20.529(0.432) & >20.8        & 19.824(0.106) & 19.480(0.092) &18.923(0.154) & fl05    \\
20170516 & 57889.89 &+120.07& 21.420(0.119) & 20.606(0.087) &  $\cdots$   &    $\cdots$     &    $\cdots$  &   $\cdots$                 & ALFOSC   \\
20170518 & 57891.15 &+121.33& $\cdots$     &   $\cdots$       &  >20.3        & 19.903(0.047) &  19.589(0.063)  &  19.279(0.120)  & ALFOSC   \\
20170530 & 57903.89 &+134.07&  $\cdots$      &    $\cdots$     &  >21.2        & 20.544(0.093) &  20.145(0.075)  &  19.767(0.204)  & ALFOSC   \\
20170618 & 57922.88 &+153.06& $\cdots$        &    $\cdots$      & $\cdots$      & 21.172(0.427) & 20.548(0.163) &20.356(0.137) & ALFOSC   \\
20170619 & 57923.89 &+154.07& $\cdots$     & 21.395(0.167) &  22.423(0.246) &  $\cdots$     & $\cdots$      & $\cdots$      & ALFOSC   \\
20170626 & 57930.89 &+161.07& $\cdots$      &   $\cdots$     &  $\cdots$      &   21.701(0.249) &   21.409(0.201)  &  20.982(0.185)  & ALFOSC   \\
20170922 & 58018.19 &+248.37& $\cdots$     &  $\cdots$      & $\cdots$      &    >22.1       &  $\cdots$       & $\cdots$      & ALFOSC   \\
20170928 & 58024.18 &+254.36& $\cdots$      &   $\cdots$     &  $\cdots$       & $\cdots$     &    >22.4        & $\cdots$       & ALFOSC   \\
20171229 & 58116.17 &+346.35&  $\cdots$       &   $\cdots$      &   >22.6          &    >22.6      &     >22.5        &       >21.2      & OSIRIS   \\
\hline
\end{tabular}                                                                                                                                            

\medskip
$^a$ Phases are relative to $r$-band maximum light, MJD = 57769.8 $\pm$ 0.1.\\ 
$^b$ The original unfiltered magnitudes reported on the Vegamag system have been rescaled to ABmags.\\
$^c$ The $u$-band magnitudes for these two epochs are reported as 19.436 (0.098) and 21.160 (0.270) mag, respectively.\\   
$^d$ The $R$ and $I$ band magnitudes for this epoch are reported as 18.037 (0.040) and 17.583 (0.035), respectively.\\  
$^e$ Lick Observatory Supernova Search (LOSS) observation (using the Katzman Automatic Imaging Telescope (KAIT; http://w.astro.berkeley.edu/bait/kait.html), Lick Observatory, California, USA) obtained through the Transient Name Server (TNS; https://wis-tns.weizmann.ac.il/object/2017be). \\
$^f$ Intermediate Palomar Transient Factory (iPTF; https://www.ptf.caltech.edu/iptf) observation using Sloan-band filters, see \citealt{2017TNSCR..43....1H}. \\
$^g$ iPTF observation (R-PTF filter, ABmag) obtained via the TNS. \\ 

\end{minipage}
\end{table*} 

\begin{table*}
\begin{minipage}{175mm}
\caption{$JHK$ photometric data for AT2017be. }
\label{JHKcurves}
\begin{tabular}{@{}ccccccl@{}}
\hline
Date & MJD & phase$^a$ & $J$(err) & $H$(err) & $K$(err) & Instrument \\
\hline
20170207 & 57791.09 & +21.27  & 16.921(0.303) & 16.102(0.269) & 15.069(0.280) & NOTCam \\
20170307 & 57819.87 & +50.05  & 17.078(0.258) & 16.423(0.227) & 15.501(0.253) & NOTCam \\
20170408 & 57851.95 & +82.13  & 17.430(0.280) & 16.508(0.376) & 15.730(0.319) & NOTCam \\
20170524 & 57897.90 & +128.08 & 18.049(0.140) & 17.265(0.089) & 16.224(0.294) & NOTCam \\
20170602 & 57906.91 & +137.09 &         --             &            --          & 16.309(0.224) &NOTCam \\
20170620 & 57924.88 & +155.06 &         --             &            --          & 16.498(0.211) & NOTCam \\
20170703 & 57937.89 & +168.07 &         --             &            --          & 16.965(0.218) & NOTCam \\
20170930 & 58026.23 & +256.41 & 19.926(0.261) & 18.881(0.307) & 17.524(0.228) & NOTCam \\
20171017 & 58043.14 & +273.32 & 19.961(0.261) & 19.346(0.395) & 17.768(0.291) & NOTCam \\
20180104 & 58122.12 & +352.30 &  >20.2         & >19.5           & 18.525(0.215) &NOTCam \\
\hline
\end{tabular}

\medskip
$^a$ Phases are relative to $r$-band maximum light, MJD = 57769.8 $\pm$ 0.1.\\ 
NOTCam: the 2.56-m Nordic Optical Telescope (NOT) with NOTCam. \\
\end{minipage}
\end{table*}

\clearpage

\label{lastpage}
\end{document}